\documentclass[aps,prl,showpacs,twocolumn,superscriptaddress]{revtex4}
\usepackage{bm,color}
\usepackage{graphicx}
\usepackage{amsmath, amssymb}
\usepackage{braket}
\usepackage{ulem}
\usepackage[dvipsnames]{xcolor}
\begin{document}
\title{Predicted Multiple Walker Breakdowns for Current-Driven Domain-Wall Motion in Antiferromagnets}
\author{Mu-Kun Lee}
\affiliation
{Department of Applied Physics, Waseda University, Okubo, Shinjuku-ku, Tokyo 169-8555, Japan}
\author{Rub\'{e}n M. Otxoa}
\affiliation
{Hitachi Cambridge Laboratory, J. J. Thomson Avenue, Cambridge CB3 OHE, UK}
\affiliation
{Donostia International Physics Center, paseo Manuel de Lardizabal 4, 20018 San Sebasti\'{a}n, Spain}
\author{Masahito Mochizuki}
\affiliation
{Department of Applied Physics, Waseda University, Okubo, Shinjuku-ku, Tokyo 169-8555, Japan}
\begin{abstract}
{We theoretically discover possible emergence of reentrant Walker breakdowns for current-driven domain walls  in layered antiferromagnets  in striking contrast to the unique Walker breakdown in ferromagnets. We reveal that the Lorentz contraction of domain-wall width in antiferromagnets gives rise to nonlinear current-dependence of the wall velocity and the predicted multiple Walker breakdowns. The dominant efficiency of the current-induced staggered spin-orbit torque over the spin-transfer torque to drive the domain-wall motion is also demonstrated. These findings are expected to be observed in synthetic antiferromagnets experimentally and provide an important contribution to the growing research field of antiferromagnetic spintronics.}
\end{abstract}
\maketitle
\textit{Introduction.}---Spintronics based on antiferromagnets has attracted significant attention in recent decades. Compared with ferromagnets, antiferromagnets possess several advantages for spintronics application, including the absence of stray fields and high-speed operation in terahertz domains~\cite{Jungwirth,Gomonay0}. Methods for manipulating and detecting spin textures in antiferromagnets including domain walls (DWs), skyrmions, bimerons, etc., have been proposed~\cite{Wadley,Zelezny,Parkin,Parkin02,Moriyama,Dohi,Akosa,Zhang,Salimath,Shen}. It is known that a moving ferromagnetic DW suffers from Walker breakdown when driven by large current or strong magnetic field, beyond which the DW oscillates between the Bloch and N\'{e}el types and its velocity is suppressed~\cite{Mougin,Yang}. Recently, it has been proposed that the antiferromagnetic DW is immune to Walker breakdown and the maximal DW speed is limited by the magnon velocity, which is, however, much higher than the breakdown threshold velocity in ferromagnets~\cite{WB_AF01,WB_AF02,Shiino,Baltz,WB_AF03}.

In this work, we theoretically study the current-driven motion of DWs in layered antiferromagnets with antiferromagnetically stacked ferromagnetic layers. We cosnider effects of both the spin-transfer torque (STT) and the staggered field-like spin-orbit torque (SOT) exerted by electric currents. We first demonstrate overwhelming efficiency of SOT over STT to drive the DW motion by numerical simulations. Then we construct an analytical theory to explain this nontrivial result and reveal the Lorentz contraction of DW as its physical origin. We further find that this DW contraction gives rise to reentrant emergence of  Walker breakdowns separated by multiple Walker regimes in which the rigid DW motion is supported. It is found that the upper limit of DW speed is still governed by magnon velocity when the Lorentz invariance manifests. Averaged DW velocities in the breakdown regimes are calculated as another prediction for future experiments. Our findings are expected to be observed in synthetic antiferromagnets~\cite{Parkin,Parkin02}.

\begin{figure}[tb]
\centering
\includegraphics[scale=1.0]{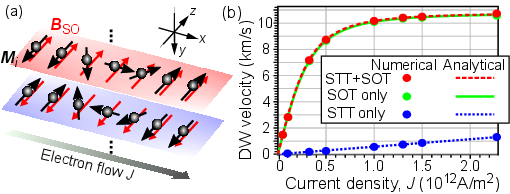}
\caption{(a) Schematics of the system. (b) Numerical and analytical results of the DW velocities.}
\label{fig_v_compare}
\end{figure}

\textit{Domain wall velocity.}---We consider antiferromagnetically stacked one-dimensional N\'{e}el DWs shown in Fig.~\ref{fig_v_compare}(a). The Hamiltonian for this system is given by,
\begin{align}
\mathcal{H}=\sum_i \Big[
&-J_{\rm F}\bm m_i \cdot \bm m_{i+\hat{x}}+J_{\rm AF}\bm m_i \cdot \bm m_{i+\hat{y}}
\nonumber\\
&+K_{\rm h}m_{iy}^2 -K_{\rm e}m_{iz}^2\Big],
\label{Hamiltonian}
\end{align}
where $\bm m_i(=\bm{M}_i/M)$ is the normalized magnetization vector at the $i$th site, $\bm M_i$ is the magnetization, and $M$ is its norm. Here $J_{\rm F}$ ($J_{\rm AF}$) is the (anti)ferromagnetic exchange coupling, and $K_{\rm e}$ ($K_{\rm h}$) is the easy (hard) magnetization anisotropy along the $z$ ($y$) axis. This Hamiltonian is a simplified model of Mn$_2$Au~\cite{Zelezny,Ruben2020,Roy} and CuMnAs~\cite{Wadley}. 

We simulate the current-induced DW motion by using the Landau-Lifshitz-Gilbert-Slonczewski (LLGS) equation~\cite{LLGS01,LLGS02,LLGS03,LLGS04}, $\partial_t \bm{M}_i=-\gamma \bm{M}_i\times [\bm B^{\rm eff}_i+(-1)^{i_y} B_{\rm SO}\hat{\bm z}] +\frac{\alpha}{M}\bm M_i \times \partial_t \bm M_i-(\bm u \cdot \bm{\nabla})\bm M_i+\frac{\beta}{M}\bm M_i \times  (\bm u \cdot \bm{\nabla})\bm M_i$.  
Here $\alpha(=0.001)$ is the Gilbert-damping coefficient, and $\beta$ is the strength of nonadiabatic torque. We introduce the current vairable $\bm u \equiv p\gamma\hbar a_0^3\bm j_{\rm e}/(2eM)$ where $\bm j_{\rm e}=j_{\rm e}\hat{\bm x}$ is the electric current density vector, $p(=0.5)$ is the spin polarization of the current, $a_0$ is the lattice constant, $\gamma(=g\mu_{\rm B}/\hbar)$ is the gyromagnetic ratio. The effective local magnetic field is calculated by $\bm B^{\rm eff}_i=-\partial \mathcal{H}/\partial\bm M_i$. The current-induced SO-field $\bm B_{\rm SO}$ alternates between the layers stacked in the $y$ direction~\cite{Zelezny,Ruben2020}. 
The parameter values are set  to be those of Mn$_2$Au (see Supplementary Information (SI) Section (Sec.)~I and II).

The injected electric current exerts both SOT and STT simultaneously to magnetizations constituting DW in each layer. Strength of the field-like SOT $B_{\rm SO}$ is proportional to the current density as $B_{\rm SO}=f_{\rm SO}j_{\rm e}$. The density-functional calculations evaluated the coefficient as $f_{\rm SO}\approx2\times 10^{-10}$ T cm$^2$/A for Mn$_2$Au~\cite{Wadley}. Here, we assume a constant $f_{\rm SO}$ for all the current-density ranges. A previous study examined the DW motion driven by SOT only~\cite{Ruben2020}. On the contrary, we investigate the current-density dependence of DW velocity in the presence of (i) only STT, (ii) only SOT, and (iii) both STT and SOT to compare the effects of SOT and STT on equal footing [Fig.~\ref{fig_v_compare}(b)]. In the simulations, we take a rather unphysically large value of $\beta$ as $\beta=10\alpha$ on purpose, with which the effect of STT should be prominent. Even in this extreme condition, the contribution of STT to DW motion is still much smaller than that of SOT. For instance, when $j_{\rm e}=1.3\times 10^{12}$~A/m$^2$, the velocity driven solely by STT ($v_{\rm STT}\approx 0.75$ km/s) is only approximately 7\% of that by SOT ($v_{\rm SOT}\approx 10.4$ km/s). This result demonstrates an overwhelming efficiency of SOT for driving the DW motion. Intriguingly, when both torques coexist, the DW velocity ($v_{\rm both}\approx 10.5$ km/s) is not given by a simple addition of these two velocities ($v_{\rm STT}+v_{\rm SOT}\approx 11.12$ km/s), although both torques should work additively~\cite{Mougin}.
To explain this phenomenon, we analytically derive the formula of DW velocity [plotted with lines in Fig.~\ref{fig_v_compare}(b)] by employing a simple two-layer model with presumed rigid DW profiles during the motion~\cite{Parkin,Mougin}. The DW profiles are obtained from the saddle-point equation of the Hamiltonian as $\bm M_l=M(\sin\theta_l\cos\phi_l, \sin\theta_l\sin\phi_l,\cos\theta_l)$ with $\theta_{\rm L}=-2\tan^{-1}[\exp\frac{x-q(t)}{\Delta}]$ and $\theta_{\rm U}=\theta_{\rm L}+\pi$, where $l$(=U, L) is an index of the upper and lower layers. Here $q(t)$ and $\Delta$ are the center coordinate and the width of the DW, respectively, and the tilt angle $\phi_l(t)$ is assumed to be spatially uniform~\cite{Parkin,Mougin}. We plugged this formula into LLGS equation and confirmed that the DWs in the upper and lower layers have common $q(t)$ and $\Delta$ when the antiferromagnetic coupling $J_{\rm AF}$ is sufficiently strong.

After some algebra, we obtain $\phi_{\rm L}=-\phi_{\rm U}$ and  $\dot\phi_{\rm U}(1+\alpha^2)=u(\alpha-\beta)/\Delta-\gamma B_{\rm SO} -\alpha\gamma (J_{\rm AF}+K_{\rm h})\sin (2\phi_{\rm U})/M$, showing a competition between STT, SOT, and antiferromagnetic exchange plus anisotropy torques. The condition for terminally static $\phi_l$, namely $\dot{\phi}_{\rm U}=0$, is
\begin{eqnarray}
\sin(2\phi_{\rm U})&=&-\frac{M}{\alpha\gamma(J_{\rm AF}+K_{\rm h})}\Big[ \frac{u(\beta-\alpha)}{\Delta }+\gamma B_{\rm SO} \Big].
~\label{sin2phiU}
\end{eqnarray}
Note that this formula has the same form as that in a ferromagnetic thin film lying on the $xy$ plane~\cite{Mougin}, where $J_{\rm AF}+K_{\rm h}$ plays the same role as the demagnetization factor $N_y-N_x$ in the latter case. Therefore, we expect Walker breakdown to occur also in the layered antiferromagnets~\cite{Ruben2020,Mougin}. Using Eq.~(\ref{sin2phiU}), the DW velocity is derived as
\begin{eqnarray}
v\equiv\dot{q}&=&u\frac{\beta}{\alpha}+\frac{\gamma B_{\rm SO}\Delta}{\alpha}.
~\label{DW_velocity}
\end{eqnarray}
The right-hand side is a sum of two contributions from STT (the first term as in~\cite{Thiaville}) and SOT (the second term). We, therefore, naively expect that the DW velocity $v$ in the presence of both STT and SOT is given by a simple sum as $v_{\rm both}=v_{\rm STT}+v_{\rm SOT}$ where $v_{\rm STT}$ ($v_{\rm SOT}$) is the velocity in the presence of STT (SOT) only. We also expect that $v$ is proportional to $u$ or the current density $j_{\rm e}$ because $B_{\rm SO}\propto u$. However, the normalized staggered magnetization $\bm{l}\equiv(\bm{M}_{\rm U}-\bm{M}_{\rm L})/2M$ in antiferromagnets follows the Lorentz-invariant equation of motion when the damping and the current-induced torques are compensated, and thus the DW width $\Delta$ suffers from a relativistic contraction~\cite{Tatara01,WB_AF01,Shiino} as $\Delta(v)\approx\Delta_0\sqrt{1-v^2/v^2_g}$, where $\Delta_0(=a_0\sqrt{J_{\rm F}/2K_{\rm e}})$ is the DW width in the static case and $v_g(=a_0\sqrt{J_{\rm F}J_{\rm AF}}/\hbar)$ is the magnon velocity in the exchange limit ($|\bm{m}|\equiv|(\bm{M}_{\rm U}+\bm{M}_{\rm L})/2M|\ll |\bm{l}|$, see SI~Sec.~IX). Therefore, $v_{\rm both}$ should depend nonlinearly on the current density $u$ and the SO-field $B_{\rm SO}$, and the velocity is no longer given by a simple addition of $v_{\rm STT}+v_{\rm SOT}$.

With the formula of the Lorentz-contracted width $\Delta(v)$, Eq.~(\ref{DW_velocity}) becomes a quadratic equation for $v$. One solution of this equation is negative and, thus, is unphysical because $\bm j_{\rm e}\propto +\hat{\bm x}$ gives rise to a net torque that should drive DW motion in the $+\hat{\bm x}$ direction (SI~Sec.~VII). The other solution is positive and thus is physical, which can be simplified as $v=u/(a\sqrt{c+du^2}-b)$ with constants $a=\gamma f_{\rm SO}/(FJ_{\rm AF})$, $b=2 K_{\rm e}\alpha \beta/F$, $c=2 J_{\rm AF}^2 J_{\rm F} K_{\rm e} (a_0\alpha)^2$, $\ d=J_{\rm AF}\hbar^2 F$, and $F=J_{\rm F} (\gamma f_{\rm SO} a_0)^2-2 K_{\rm e}\beta^2$. This simple analytical solution is the first major result of this study, which is shown by lines in Fig.~\ref{fig_v_compare}(b) and coincides well with the numerical results.
We show an alternative derivation based on the Thiele equation~\cite{Thiele} in SI~Sec.~X, which turns out to give the same formula for the DW velocity.

When $B_{\rm SO}=0$, $v$ is independent of $J_{\rm AF}$, which can be understood intuitively. A finite $B_{\rm SO}$ efficiently tilts $\bm{M}_l$ on both layers along the hard axis $-\hat{\bm y}$ direction (SI~Sec.~VII). Subsequently, the $J_{\rm AF}$ coupling induces a strong exchange torque owing to this tilt to drive DW motion~\cite{Roy}. This scenario is the same as that in the synthetic antiferromagnets~\cite{Parkin}, where the damping-like SOT is the dominant mechanism for the high DW speed. Note that Eq.~(\ref{sin2phiU}) indicates that there is still a finite tilt angle even when $B_{\rm SO}=0$ owing to the STT. 
The DW velocity in this case is proportional to $(J_{\rm AF}+K_{\rm h})\Delta\sin(2\phi_{\rm U})$ (see SI~Sec.~II), and thus, the dependence on $J_{\rm AF}$ is cancelled, resulting in effectively uncoupled ferromagnetic DWs. In addition, in the limit of $u\rightarrow \infty$, we obtain $v\rightarrow v_g\sqrt{1-n}$, with $n=\frac{2K_{\rm e}}{J_{\rm F}}(\frac{\beta}{\gamma f_{\rm SO}a_0})^2\approx 4\times 10^{-4}$ using $\beta=10\alpha$. Therefore, we find that the DW velocity cannot exceed the magnon velocity $v_g$ as expected and can reach $v_g$ only in the adiabatic limit of $\beta=0$.

\textit{Magnon velocity and tilt angle.}---We use a saddle-point solution of $\theta_{\rm U,L}$ to fit the simulated DW widths and evaluate $v_g$ numerically, which is close to the analytical value (SI~Sec.~III). For $j_{\rm e}=1.5\times 10^{12}$~A/m$^2$, Eq.~(\ref{sin2phiU}) leads to $\sin(2\phi_{\rm U})\approx-0.16$, and the DWs are within Walker regime to maintain the rigid moving profiles, which self-consistently justifies our initial substitution of the rigid DW profiles into the LLGS equation. This corresponds to $M_{\text{U/L},y}\approx -0.08M\text{sech}(\frac{x-vt}{\Delta(v)})$ with the same sign for both layers. This analytical result coincides well with the simulation results (SI~Sec.~III).  The minor discrepancy in the peak height is attributable to nonzero spin-wave emissions behind the moving DW indeed observed in the simulations. 

\textit{Prediction of multiple Walker regimes.}---
The Walker breakdown is defined as a regime in which rigid DW profiles are no longer stable when driven by large current or strong field exceeding a threshold, with which $\dot{\phi}_{\rm U,L}$ becomes finite. In the breakdown regime, the right-hand side of Eq.~(\ref{sin2phiU}) is greater than $1$ or less than $-1$, and thus the threshold current $u_{\rm c}$ is determined by the condition $\sin(2\phi_{\rm U}(u_{\rm c}))=\pm 1$. It has been widely believed that the Walker breakdown should not occur for DWs in antiferromagnets~\cite{WB_AF01,WB_AF02,Shiino,Baltz,WB_AF03}. However, we find that it can occur in layered antiferromagnets with exchange coupling $J_{\rm AF}$ because Eqs.~(\ref{sin2phiU}) and (\ref{DW_velocity}) have the same form as that for DWs in ferromagnetic thin films~\cite{Mougin}. Substituting the analytical solution of $v$ into Eq.~(\ref{sin2phiU}), we obtain its $u$-dependence as $\sin(2\phi_{\rm U})=-c_0u\Big[1+\frac{(\beta-\alpha)(c_1+c_2 u^2)}{x_1\sqrt{1+c_5 u^2}-x_2 u^2}\Big]$ with coefficients specified in SI~Sec.~IV. This indicates nonlinear $u$ dependence of $\sin(2\phi_{\rm U})$ and possible emergence of multiple Walker regimes separated by breakdown regimes with boundaries defined by $\sin(2\phi_{\rm U}(u_{\rm c}))=\pm 1$. This is in striking contrast to the case of ferromagnets, in which  $\sin(2\phi)$ shows a monotonic behavior against $u$ and, thus, only a unique threshold current density appears~\cite{Mougin,Yang}.

\begin{figure}[tb]
\centering
\includegraphics[scale=1.3]{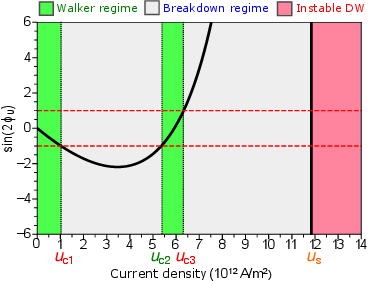}
\caption{Thick solid (black) curves show $\sin(2\phi_{\rm U})$ when $\beta=0.5\alpha$. Horizontal dotted lines label the values $\pm 1$. Grey areas indicate the Walker breakdown regimes, and red area the instable DW regime.}
\label{sin_fig}
\end{figure}

For better data visualization, Fig.~\ref{sin_fig} shows the current-density dependence of $\sin(2\phi_{\rm U})$ when $\alpha=0.005,\beta=0.5\alpha$, and $J_{\rm AF}=10^{-3}J_{\text{AF,Mn$_2$Au}}$ with $J_{\text{AF,Mn$_2$Au}}$ being the AF exchange coupling in Mn$_2$Au, in which we find multiple Walker regimes in (i) $0<u<u_{\rm c1}$ with $u_{\rm c1}\approx u_0$ and (ii) $u_{\rm c2}< u< u_{\rm c3}$ with $u_{\rm c2}\approx 5.4 u_0$ and $u_{\rm c3}\approx 6.3 u_0$, where $u_0\equiv (p\gamma\hbar a_0^3/2eM) \times 10^{12}$ A/m$^2$. There is a singular point of $\sin(2\phi_{\rm U})$ at $u_{\rm s}=v_{g}\alpha/\beta\approx 11.8u_0$, at which the denominator vanishes as $x_1\sqrt{1+c_5u^2}-x_2u^2=0$ that causes $\sin(2\phi_{\rm U})$ abruptly crosses from a positive to a negative value. This $u_{\rm s}$ is a criterion of $u$ for the stability of DW. Equation~(\ref{DW_velocity}) leads to $\Delta=(\alpha v-\beta u)/\gamma f_{\rm SO}u$. At the singular current, $\Delta(u_{\rm s})=(\alpha v(u_{\rm s})-\beta u_{\rm s})/\gamma f_{\rm SO}u_{\rm s}\le  (\alpha v_g-\beta u_{\rm s})/\gamma f_{\rm SO}u_{\rm s}=0$, thus the DW already shrinks to zero width before the current reaches $u_{\rm s}$. Therefore, for current $u>u_{\rm s}$, even if there is a self-consistent Walker regime, it does not support a stable DW. The equation for the threshold current, $\sin(2\phi_{\rm U}(u_{\rm c}))=\pm 1$, has no analytical solutions since it has a form $\sum^6_{n=1}\gamma_n u^n_{\rm c}=0$ with the sixth-order polynomial of $u_{\rm c}$ and coefficients $\gamma_n$. Nevertheless, it evidently implies that  more than one Walker regimes can exist. This is another central result of this work. Note that there is also self-consistent solution of multiple Walker breakdowns when $\beta>\alpha$ (SI~Sec.~IV), but it cannot be observed since the second Walker regime occurs at currents larger than $u_{\rm s}$. For $\alpha=\beta$, only a single breakdown regime appears similar to the ferromagnetic case, which leads to $\sin(2\phi_{\rm U})=-c_0u$ and a unique threshold current density of $u_{\rm c}=1/c_0$.

The physical mechanism of reentrant Walker regime can be seen by the equation of $\dot{\phi}_{\rm U}$ in the paragraph above Eq.~(\ref{sin2phiU}). The strength of STT due to angular momentum conservation when electrons pass through the DW is proportional to $u/\Delta$ which characterizes the rate of electron spin reverse, since $\Delta$ is the length scale of local magnetization reverse in a DW, and $u$ relates to polarized electron's velocity. Due to the Lorentz contraction of $\Delta$ in antiferromagnets, the competition between STT ($\propto u/\Delta$) and SOT ($\propto u$) can bend the right-hand side of Eq.~(\ref{sin2phiU}) from less than $-1$ to exactly $-1$, driving the DW from breakdown into second Walker regime, as confirmed by numerical integration of LLGS equation with fourth-order Runge-Kutta (RK) method [see SI~Sec.~V].

\begin{figure}[tb]
\centering
\includegraphics[scale=0.5]{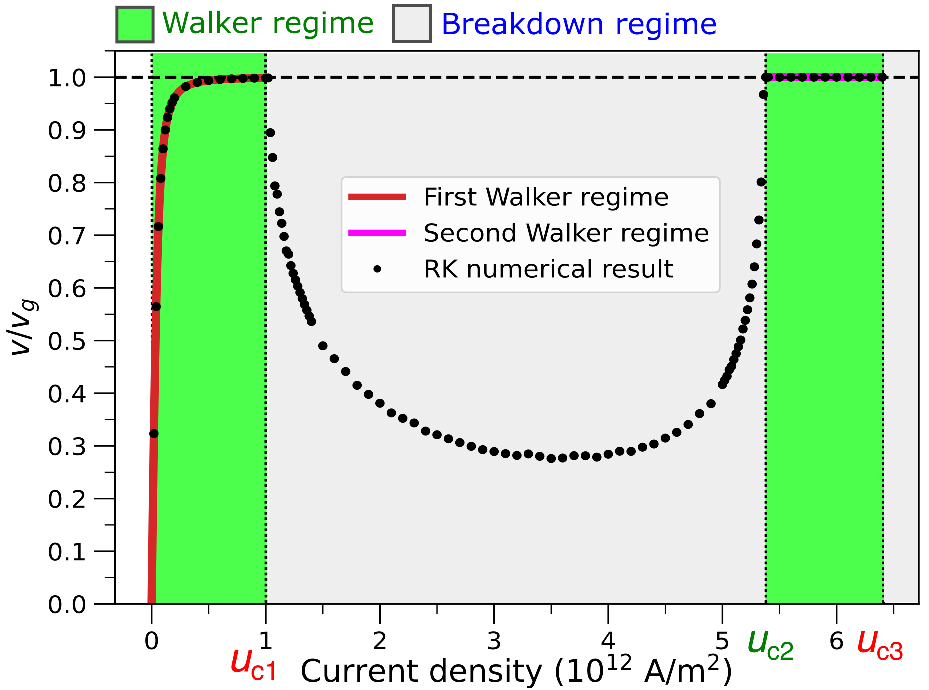}
\caption{Curves represent analytical DW velocities in Walker regimes, and dots represent terminal (time-averaged) velocities in Walker (breakdown) regimes by RK calculation.}
\label{fig_v_vs_u}
\end{figure}

\textit{Averaged velocity in breakdown regime.}---In breakdown regime, $\phi_{\rm U,L}(t)$ depends on time such that both $\sin(2\phi_{\rm U, L}(t))$ and $v(t)$ oscillate in time, and it is no longer permissible to use Eq.~(\ref{sin2phiU}) to obtain Eq.~(\ref{DW_velocity}). From RK calculation (SI~Sec.~VI), we get the time-averaged DW velocity as shown in black dots in Fig.~\ref{fig_v_vs_u}. Besides a quantitative agreement with the analytical velocities in Walker regimes, in breakdown regime we find a drop of time-averaged velocity similar to the case in ferromagnets~\cite{Mougin,Yang}.

There is an intuitive way to expect this velocity drop in breakdown regime. In the presence of STT and staggered SOT, we can extend the consideration in~\cite{Dasgupta,Belashchenko,Saarikoski} to write the Lagrangian density for our layered antiferromagnet as,
\begin{eqnarray}
\mathcal{L}=-\mathcal{J}\bm{m}\cdot[\bm{l}\times (\partial_t+u\partial_x)\bm{l}]-4J_{\rm AF}\bm{m}^2/a_0-\mathcal{U}(\bm{l}),~\label{Lagrangian}
\end{eqnarray}
where $\mathcal{J}=M/\gamma a_0$ and $\mathcal{U}=J_{\rm F}a_0(\partial_x\bm{l})^2-K_{\rm e}l^2_z/a_0+K_{\rm h}l^2_y/a_0-(M/a_0)\bm{l}\cdot\bm{B}_{\rm SO}$ is the energy density of $\bm{l}$. The $\mathcal{J}$ term is the spin Berry phase in a gauge with opposite Dirac strings $\bm{n}_{0,l}$ for the two sublattices in the monopole representation $\sum_l\mathcal{L}_{\text{B},l}=\frac{1}{\gamma a_0}\sum_l[\bm{n}_{0,l}\cdot\bm{M}_l\times (\partial_t+u\partial_x)\bm{M}_l]/(1-\bm{n}_{0,l}\cdot\bm{M}_l)$ and is expanded up to the second order of $\bm{m}$. (The next finite order of $\bm{m}$ is the third order which can be proved by choosing, e.g., $\bm{n}_{0}=\pm\hat{\bm{z}}$ for the two sublattices, respectively.)~\cite{Dasgupta,Belashchenko}. The adiabatic STT contributes to the second term of the convective derivative $(\partial_t+u\partial_x)$~\cite{Tatara2008}. Neglecting the nonadiabatic STT and Rayleigh dissipation ($\propto\alpha\dot{\bm{m}}^2$), we obtain $\bm{m}=-\mathcal{J}a_0\bm{l}\times(\partial_t+u\partial_x)\bm{l}/(8J_{\rm AF})$ from the Lagrange equation. Using the saddle-point DW profile of $\bm{l}$, the $\bm{m}$-dependent terms after integrating over $x$ become $\frac{M_0}{2}(v-u)^2+\frac{I}{2}\dot{\phi}_{\rm U}^2$, which are the translational and rotational kinetic energies of a soliton with mass $M_0=M^2/(4\gamma^2a_0\Delta J_{\rm AF})$ and moment of inertia $I=M_0\Delta^2$. In this derivation, we assume a time-independent terminal $\Delta$.

Near $u_{\rm c1}\approx 0.06$ km/s, $v$ is close to $v_g\approx 0.34$ km/s, thus $v(u_{\rm c1})$ is greater  than $u_{\rm c1}$. When the system crosses the threshold $u_{\rm c1}$ and enters the breakdown regime, the soliton angular frequency $\dot\phi_{\rm U}$ changes abruptly from zero to finite, which causes a nonzero rotational energy. To preserve the kinetic energy of the soliton in a narrow current-density range near $u_{\rm c1}$, the DW velocity should decrease as Fig.~\ref{fig_v_vs_u} shows. 
From the experimental point of view, however, it should be mentioned that with currents close to $u_{\rm c1}$, several other instabilities and effects may appear such as spin-wave emissions~\cite{Tatara_SWemit} and DW proliferations~\cite{Ruben2020}~(SI.~Sec.~VII). Indeed, it was theoretically argued that an effective gyrofield induced by the kinetic energies of DW can cause the DW proliferation together with the transient Lorentz-invariance breaking with DW speeds exceeding $v_g$~\cite{Ruben2020,Guslienko}.
\begin{figure}[tb]
\centering
\includegraphics[scale=0.47]{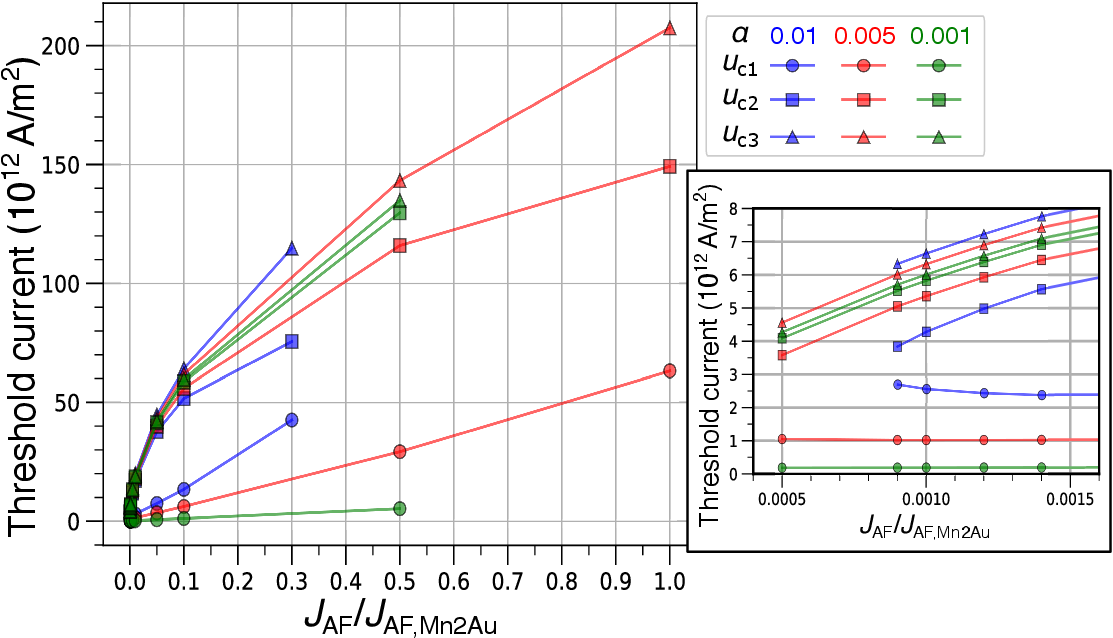}
\caption{Threshold currents as functions of the interlayer antiferromagnetic coupling for various Gilbert dampings. Inset shows enlarged view.}
\label{fig_uc}
\end{figure}

\textit{Proposal for experimental observations.}---In synthetic antiferromagnets~\cite{Parkin,Parkin02}, an underlying Pt layer induces a damping-like SOT $\propto \bm{M}\times \bm{M}\times \hat{\bm{z}}$ (using our coordinate convention) in the same direction but with different strengths for the two magnetic layers on top of it (stacked along $-\hat{\bm{y}}$)~\cite{Parkin}. One can fabricate another Pt layer above the upper magnetic layer which, by symmetry, generates an opposite SOT when applying the current. In this situation, each layer is driven by opposite damping-like SOT. After taking a curl product of the LLGS equation with $\bm{M}$ to cancel time derivatives on its right-hand side, it induces a staggered field-like SOT $\propto \pm\bm{M}\times(\bm{M}\times\bm{M}\times\hat{\bm{z}})\propto\mp\bm{M}\times\hat{\bm{z}}$, which mimics the staggered SO-field $\bm B_{\rm SO}$ in our case. Therefore, we expect similar results of multiple Walker breakdowns to occur in synthetic antiferromagnets with both top and bottom Pt layers. 

Since $J_{\rm AF}$ can be tuned in synthetic antiferromagnets by changing the thickness of metallic layer sandwiched by two magnetic layers, we investigate the dependence of $u_{\text{c}j}$ on $J_{\rm AF}$ as well as the Gilbert damping $\alpha$ when $\beta=0.5\alpha$ with other parameters following that of Mn$_2$Au. The results plotted in Fig.~\ref{fig_uc} show nearly monotonic decrease of $u_{\text{c}j}$ as $J_{\rm AF}$ decreases, since a smaller antiferromagnetic exchange torque requires a smaller STT (and thus smaller $u_{\text{c}j}$) to compensate as shown in the equation for $\dot{\phi}_{\rm U}$ above Eq.~(\ref{sin2phiU}). It requires roughly $J_{\rm AF}\le 10^{-3}J_{\text{AF,Mn$_2$Au}}$ to get experimentally feasible current below $\sim10^{13}$~A/m$^2$ to observe the second Walker regime.
We have checked the Lorentz contraction of DW width still manifests in this small exchange regime by micromagnetic simulation. Moreover, for larger dampings, second Walker regime between $u_{\rm c2}$ and $u_{\rm c3}$ appears in a wider range, appropriate for experimental observations, although one should note that for $\alpha=0.01$, there is no longer real solution of $u_{\text{c}j}$ for $J_{\rm AF}<9\times10^{-4}J_{\text{AF,Mn$_2$Au}}$ in our parameter set.

\textit{Conclusion.}---We have theoretically studied the DW motion in layered antiferromagnets driven by electric current, which exerts both STT and staggered SOT. We have discovered possible reentrant emergence of multiple Walker breakdowns, which is in sharp contrast to the unique Walker breakdown for the current-driven DW motion in ferromagnets. We have revealed that the Lorentz contraction of DW width in antiferromagnets gives rise to nonlinear current-dependence of the DW velocity and the predicted multiple Walker breakdowns. The dominant efficiency of SOT over STT and their non-additive effects in driving the DW motion have been also demonstrated. It should be mentioned that the present theory can be generalized in the straightforward way for intrinsic antiferromagnetic systems in which STT is not applicable or for the cases in which other torques due to additional effects are present (e.g., Dzyaloshinskii-Moriya interaction, Rashba spin-orbit interaction, spin Hall effect). Our findings are expected to be observed in synthetic antiferromagnets experimentally and provide siginificant contributions to development of the antiferromagnetic spintronics.

This work is supported by Japan Society for the Promotion of Science KAKENHI (Grant No. 20H00337 and No. 23H04522), CREST, the Japan Science and Technology Agency (Grant No. JPMJCR20T1) , and the Waseda University Grant for Special Research Project (Project No. 2023C-140). M.K.L. is grateful for illuminating discussions with Rintaro Eto, Collins A. Akosa, and Xichao Zhang.


\begin{thebibliography}{999}
\bibitem{Jungwirth}T. Jungwirth, X. Marti, P. Wadley, and J. Wunderlich, Nat. Nanotech. \textbf{11}, 231 (2016).
\bibitem{Gomonay0}E. V. Gomonay and V. M. Loktev, Low Temp. Phys. \textbf{40}, 17 (2014).
\bibitem{Wadley}P. Wadley, B. Howells, J. Zlezn\'{y}, C. Andrews, V. Hills, R. P. Campion, V. Nov\'{a}k, K. Olejn\'{i}k, F. Maccherozzi, S. S. Dhesi, S. Y. Martin,  T. Wagner, J. Wunderlich, F. Freimuth, Y. Mokrousov, J. Kune\v{s}, J. S. Chauhan, M. J. Grzybowski, A. W. Rushforth, K. W. Edmonds, B. L. Gallagher, T. Jungwirth, Science \textbf{351}, 587 (2016).
\bibitem{Zelezny}J. \v{Z}elezn\'{y}, H. Gao, K. V\'yborn\'{y}, J. Zemen, J. Ma\v{s}ek, A. Manchon, J. Wunderlich, J. Sinova, and T. Jungwirth, Phys. Rev. Lett. \textbf{113}, 157201 (2014).
\bibitem{Parkin}S. H. Yang, K. S. Ryu, and S. Parkin, Nat. Nanotech. \textbf{10}, 221 (2015).
\bibitem{Parkin02}R. A. Duine, K.-J. Lee, S. S. P. Parkin, and M. D. Stiles, Nat. Phys. \textbf{14}, 217 (2018).
\bibitem{Moriyama}T. Moriyama, W. Zhou, T. Seki, K. Takanashi, and T. Ono, Phys. Rev. Lett. \textbf{121} 167202 (2018).
\bibitem{Dohi}T. Dohi, S. DuttaGupta, S. Fukami, and H. Ohno, Nat. Comm. \textbf{10} 5153 (2019).
\bibitem{Akosa}C. A. Akosa, O. A. Tretiakov, G. Tatara, A. Manchon, Phys. Rev. Lett. \textbf{121} 097204 (2018).
\bibitem{Zhang}X. Zhang, Y. Zhou, and M. Ezawa, Sci. Rep. \textbf{6}, 24795 (2016).
\bibitem{Salimath}A. Salimath, Fengjun Zhuo, R. Tomasello, G. Finocchio, and A. Manchon, Phys. Rev. B \textbf{101}, 024429 (2020).
\bibitem{Shen}L. Shen, J. Xia, X. Zhang, M. Ezawa, O. A. Tretiakov, X. Liu, G. Zhao, and Y. Zhou, Phys. Rev. Lett. \textbf{124}, 037202 (2020).
\bibitem{Mougin}A. Mougin, M. Cormier, J. P. Adam, P. J. Metaxas, and J. Ferr\'{e}, Europhys. Lett. \textbf{78}, 57007 (2007).
\bibitem{Yang}J. Yang, C. Nistor, G. S. D. Beach, and J. L. Erskine, Phys. Rev. B \textbf{77}, 014413 (2008).
\bibitem{WB_AF01}O. Gomonay, T. Jungwirth, and J. Sinova, Phys. Rev. Lett. \textbf{117}, 017202 (2016).
\bibitem{WB_AF02}S. Selzer, U. Atxitia, U. Ritzmann, D. Hinzke, and U. Nowak, Phys. Rev. Lett. \textbf{117}, 107201 (2016).
\bibitem{Shiino}T. Shiino, S.-H. Oh, P. M. Haney, S.-W. Lee, G. Go, B.-G. Park, and K.-J. Lee, Phys. Rev. Lett. \textbf{117}, 087203 (2016).
\bibitem{Baltz}V. Baltz, A. Manchon, M. Tsoi, T. Moriyama, T. Ono, and Y. Tserkovnyak, Rev. Mod. Phys. \textbf{90}, 015005 (2018).
\bibitem{WB_AF03}O. Gomonay, T. Jungwirth, and J. Sinova, physica status solidi (RRL), Rapid Research Letters \textbf{11}, 1700022 (2017).
\bibitem{Ruben2020}R. M. Otxoa, P. E. Roy, R. Rama-Eiroa, J. Godinho, K. Y. Guslienko, and J. Wunderlich, Commun. Phys. \textbf{3}, 190 (2020).
\bibitem{Roy}P. E. Roy, R. M. Otxoa, and J. Wunderlich, Phys. Rev. B \textbf{94}, 014439 (2016).
\bibitem{LLGS01}L. D. Landau and E. M. Lifshitz, Phys. Z. Sowjetunion \textbf{8} 153 (1935).
\bibitem{LLGS02}T. L. Gilbert, Phys. Rev. \textbf{100} 1243 (1955).
\bibitem{LLGS03}S. Zhang and Z. Li, Phys. Rev. Lett. \textbf{93} 127204 (2004).
\bibitem{LLGS04}G. Tatara and H. Kohno, Phys. Rev. Lett. \textbf{92} 086601 (2004).
\bibitem{Thiaville}A. Thiaville, Y. Nakatani, J. Miltat, and Y. Suzuki, Europhys. Lett. \textbf{69}, 990 (2005).
\bibitem{Tatara01}G. Tatara, C. A. Akosa, and R. M. Otxoa de Zuazola, Phys. Rev. Research \textbf{2}, 043226 (2020).
\bibitem{Thiele}A. A. Thiele, Phys. Rev. Lett. \textbf{30}, 230 (1973).
\bibitem{Dasgupta}S. Dasgupta, S. K. Kim, and O. Tchernyshyov, Physical Review B \textbf{95}, 220407 (2017).
\bibitem{Belashchenko}K. D. Belashchenko, O. Tchernyshyov, A. A. Kovalev, and O. A. Tretiakov, Applied Physics Letters \textbf{108}, 132403 (2016).
\bibitem{Saarikoski}H. Saarikoski, H. Kohno, C. H. Marrows, G. Tatara, Phys. Rev. B \textbf{90}, 094411 (2014). 
\bibitem{Tatara2008}G. Tatara, H. Kohno, J. Shibata, Phys. Rep. \textbf{468}, 213 (2008).
\bibitem{Tatara_SWemit}G. Tatara and R. M. Otxoa de Zuazola, Phys. Rev. B \textbf{101}, 224425 (2020).
\bibitem{Guslienko}K. Y. Guslienko, K. S. Lee, and S. K. Kim, Phys. Rev. Lett. \textbf{100}, 027203 (2008).
\end{thebibliography}
\end{document}


{\centering{\large\textbf{Supplementary Information: \\Predicted Multiple Walker Breakdowns for Current-Driven Domain-Wall Motion in Antiferromagnets}}}\\ \\
Mu-Kun Lee$^1$, Rub\'{e}n M. Otxoa$^{2,3}$, and Masahito Mochizuki$^1$\\
{\footnotesize \textit{$^1$Department of Applied Physics, Waseda University, Okubo, Shinjuku-ku, Tokyo 169-8555, Japan\\
$^2$Hitachi Cambridge Laboratory, J. J. Thomson Avenue, Cambridge CB3 OHE, UK\\
$^3$Donostia International Physics Center, paseo Manuel de Lardizabal 4, 20018 San Sebasti\'{a}n, Spain}}
\tableofcontents
\section{Details of micromagnetic simulation}
For micromagnetic simulations, we take a system containing one-dimensional chains each with 60000 atomic sites along $x$-direction as shown in Fig.~1~(a) in the main text. The open boundary condition (BC) is used for $x$- and $z$-directions, while periodic BC is used for the $y$-direction. Initial domain wall (DW) configuration is obtained by relaxing for a sufficiently long time a presumed saddle-point antiferromagnetically-coupled DW profile close to the left edge of the chains using the static DW width $\Delta_0=a_0\sqrt{J_{\rm F}/2K_{\rm e}}\approx 25a_0$, where $J_{\rm F}$, $K_{\rm e}$, and $a_0$ are the ferromagnetic exchange constant, easy-axis anisotropy, and lattice constant, respectively, with magnitudes being specified in the next section. The electric current is applied along the $x$-direction, with a linear ramp-up time as 100 ps from zero to the full magnitude to reduce spin wave emissions or other transient effects. The LLGS equation for magnetization dynamics under current is numerically solved by using the fourth-order Runge-Kutta (RK) method. DW center as a function of time is measured by the site with the maximal $M_x$ component in time. We have checked in our current model parameters (see next section) and range of currents that each DWs have the same center during motion. DW velocity is calculated by subtracting the DW center at 1000 ps and that at 900 ps and dividing it by the time difference of 100 ps. We have checked the velocities in this time range have already saturated.
\section{Analytical calculation of DW velocity}
We consider a system of two antiferromagnetically (AF) coupled one-dimensional domain walls (DWs), with their magnetizations in the discrete lattice written as $\bm{m}_{l,i}=\bm{M}_{l,i}/M=(\sin\theta_{l,i}\cos\phi_{l,i}, \sin\theta_{l,i}\sin\phi_{l,i}, \cos\theta_{l,i})$, where subscript $l=U(L)$ stands for the upper (lower) layer, subscript $i$ is the site index along $x$ direction, and $M(=4\mu_{\rm B})$ is the magnetization ($\mu_{\rm B}$ stands for the Bohr magneton). Define the magnetization density vector in the continuous limit as $\bm{m}_l(x)=\bm{m}_{l,i}/a_0$ (and $\bm{M}_l(x)=\bm{M}_{l,i}/a_0, \theta_l(x)=\theta_{l,i}, \phi_l(x)=\phi_{l,i}$) with lattice constant $a_0=0.3328$~nm in the following. We consider the Hamiltonian density along $\hat{\bm{x}}$ direction, $\mathcal{H}(x,t)$, as
\begin{eqnarray}
\mathcal{H}(x,t)&=&\sum_{l=\rm U,L}\Bigg\{\frac{J_{\rm F}a_0}{2}\Big[(\partial_x\theta_l)^2+\sin^2\theta_l(\partial_x\phi_i)^2\Big]+\frac{1}{a_0}\Big[K_{\rm h}\sin^2\theta_l\sin^2\phi_l - K_{\rm e}\cos^2\theta_l\Big]\Bigg\}\nonumber\\
&+&\frac{B_{\rm SO}M}{a_0}\Big[\cos\theta_{\rm U}-\cos\theta_{\rm L}\Big]+\frac{J_{\rm AF}}{a_0} \Big[\sin\theta_{\rm L}\sin\theta_{\rm U}\cos(\phi_{\rm L}-\phi_{\rm U})+\cos\theta_{\rm L}\cos\theta_{\rm U}\Big]~\label{H}
\end{eqnarray}
where $J_{\rm F}(=9.91\times 10^{-3}$ eV) and $J_{\rm AF}(=4.58\times 10^{-2}$ eV) are the ferromagnetic and AF exchange constant along $\hat{\bm{x}}$ and $\hat{\bm{y}}$ direction, respectively. $K_{\rm h}(=8.1\times 10^{-4}$ eV) is hard-axis anisotropy along $\hat{\bm{y}}$, and $K_{\rm e}(=8.1\times 10^{-6}$ eV) is easy-axis anisotropy along $\hat{\bm{z}}$. All model parameters are referred from the values in Mn$_2$Au~\cite{Ruben2020}.

The effective field is calculated by $\bm{B}^{\rm eff}_l(x,t)=\frac{-1}{a_0}\delta[\int dx' \mathcal{H}(x')]/\delta\bm{M}_l(x,t)$. Plugging the presumed DW solution $\theta_{\rm L}=-2\tan^{-1}[\exp\frac{x-q_{\rm L}(t)}{\Delta(\dot{q}_{\rm L})}]$ and $\theta_{\rm U}=-2\tan^{-1}[\exp\frac{x-q_{\rm U}(t)}{\Delta(\dot{q}_{\rm U})}]+\pi$ into the LLGS equation as shown in the main text, we can derive the equations of motion for DW velocity $\dot{q}_i$ and hard-axis tilt angle $\phi_{i}$. To this end, we assume (i) both DWs in upper and lower chains have the same center during motion, $q_{\rm U,L}(t)=q(t)$, which has been confirmed by simulation within our model parameters and excitation protocol; (ii) since $\phi_{\rm U,L}(t)$ is the conjugate momentum of $q_{\rm U,L}(t)$ when the Lagrangian density contains only the spin Berry phase, $\dot{\phi_l}(\cos\theta_l-1)$, it should only depends on time but not on space in that case~\cite{Tatara2014}.
Here we also assume $\phi_{\rm U,L}(t)$ only depends on time but not on $x$ and take $\partial_x\phi_{\rm U,L}=0$ as in~\cite{Parkin}. This is a rather strong assumption, since the Hamiltonian density as Eq.~(\ref{H}), taking the DW profile, includes interaction terms between $q_l$ and $\phi_l$~\cite{Tatara2014}, thus $\phi_l$ is no longer strictly the conjugate momentum of $q_l$ and there is no guarantee that $\phi_l$ should not depend on space. Nevertheless, this assumption works well in several works when we focus on the motion of DW center~\cite{Parkin,Mougin}. Lastly, we assume (iii) the relativistic DW width $\Delta(\dot{q})=\sqrt{\frac{J_{\rm F}a^2_0}{2K_{\rm e}}(1-\frac{\dot{q}^2}{v^2_{g}})}$ is independent of time; namely, we take the terminal velocity limit in which $\dot{q}$ approaches a constant $v$. In this case, $\partial\bm{M}_{l,i}/\partial t$ in LLGS equation only contains time derivatives of $q$, but not of $\Delta$.

Although in assumption (iii) we approximate DW width $\Delta$ as being time-independent, first let us consider the general case when $\Delta = \Delta(t)$ also depends on time. Substituting the DW profiles into LLGS equation in the continuous limit, we get [denote ``()" as ``$(\frac{x-q}{\Delta})$"]
\begin{eqnarray}
&&\dot{M}_{l,x}=\frac{M}{a_0\Delta}\text{sech}()\text{tanh}()(\pm\cos\phi_i)\Big(\dot{q}_l+\frac{x-q_l}{\Delta}\dot{\Delta}\Big)+\frac{M}{a_0}\text{sech}()(\mp\sin\phi_l)\dot{\phi}_l=\tau_{l,x},\label{dMux_dt}\nonumber\\
&&\dot{M}_{l,y}=\frac{M}{a_0\Delta}\text{sech}()\text{tanh}()(\pm\sin\phi_l)\Big(\dot{q}_l+\frac{x-q_l}{\Delta}\dot{\Delta}\Big)+\frac{M}{a_0}\text{sech}()(\pm\cos\phi_l)\dot{\phi}_l=\tau_{l,y}\label{dMuy_dt},\nonumber\\
&&\dot{M}_{l,z}=\mp\frac{M}{a_0\Delta}\text{sech}^2()\Big(\dot{q}_l+\frac{x-q_l}{\Delta}\dot{\Delta}\Big)=\tau_{l,z}\label{dMuz_dt},\nonumber
\end{eqnarray}
where we write $\bm{\tau}_{l}$ as the torque and the upper (lower) sign is for upper (lower) layer. From these equations we get
\begin{eqnarray}
\dot{q}_{\text{U/L}}&+&\frac{x-q_{\text{U/L}}}{\Delta}\dot{\Delta}=\frac{\pm a_0\Delta}{M}\frac{(\cos\phi_{\text{U/L}})\tau_{{\text{U/L}},x}+(\sin\phi_{\text{U/L}})\tau_{{\text{U/L}},y}}{\text{sech()tanh()}}=\frac{\mp a_0\Delta \tau_{{\text{U/L}},z}}{M\text{sech}^2()},\label{qu_eq}\\
\dot{\phi}_{\text{U/L}}&=&\frac{a_0}{M}\frac{\mp (\sin\phi_{\text{U/L}})\tau_{{\text{U/L}},x}\pm(\cos\phi_{\text{U/L}})\tau_{{\text{U/L}},y}}{\text{sech}()}.
\end{eqnarray}
Substituting all the torques in LLGS equation $\bm{\tau}_l$ derived from the Hamiltonian and spin transfer torques into right-hand sides, we find from above equations
\begin{eqnarray}
\dot{q}_{\rm U,L}&=&\frac{u(1+\alpha\beta)}{1+\alpha^2}+\frac{B_{\rm SO}\alpha \gamma \Delta}{(1+\alpha^2)}+\frac{J_{\rm AF} \gamma \Delta \sin(\phi_{\rm L}-\phi_{\rm U})}{M(1+\alpha^2)}\mp\frac{K_{\rm h}\gamma \Delta \sin (2\phi_{\rm U,L})}{M(1+\alpha^2)}-\Big(\frac{x - q_{\rm U,L}}{\Delta}\Big)\dot{\Delta}\label{qU_dot_Delta}\\
&-&\text{tanh}\Big(\frac{x-q_{\rm U,L}}{\Delta}\Big)\frac{\alpha\gamma }{M(1+\alpha^2)}\Big[\Delta\Big(J_{\rm AF}+2K_{\rm e}+K_{\rm h}-J_{\rm AF}\cos(\phi_{\rm L}-\phi_{\rm U})-K_{\rm h}\cos (2\phi_{\rm U,L})\Big)-\frac{J_{\rm F}a^2_0}{\Delta}\Big]\label{qU_fac},\\
\dot\phi_{\rm U,L}&=&\pm\frac{ u(\alpha-\beta)}{\Delta (1+\alpha^2)}\mp\frac{B_{\rm SO}\gamma }{(1+\alpha^2)}\pm\frac{J_{\rm AF} \alpha\gamma \sin(\phi_{\rm L}-\phi_{\rm U})}{M(1+\alpha^2)}-\frac{K_{\rm h}\alpha\gamma\sin (2\phi_{\rm U,L})}{M(1+\alpha^2)}~\label{dtphiU}\\
&\pm&\text{tanh}\Big(\frac{x-q_{\rm U,L}}{\Delta}\Big)\frac{\gamma }{M (1+\alpha^2)}\Big[\Big(J_{\rm AF}+2K_{\rm e}+K_{\rm h}-J_{\rm AF}\cos(\phi_{\rm L}-\phi_{\rm U})-K_{\rm h}\cos( 2\phi_{\rm U,L})\Big)-\frac{J_{\rm F}a^2_0}{\Delta^2}\Big],
\end{eqnarray}
where the upper (lower) signs are for the upper (lower) layer. In the end of this section we will show that we need to set $x=q_l(t)$ on all right-hand sides such that all $\text{tanh}(\frac{x-q_{l}}{\Delta})$ terms are zero, and remaining terms are independent of $x$. This is equivalent to focusing on the DW center, an approximate approach commonly adopted when there are external current or magnetic fields which make the formula of $\theta_{l}$ no longer the exact solution of the LLGS equation~\cite{Parkin,Mougin}. For now we just take this procedure and get the following equations
\begin{eqnarray}
\dot{q}_{\rm U,L}&=&\frac{u(1+\alpha\beta)}{1+\alpha^2}+\frac{B_{\rm SO}\alpha \gamma \Delta}{(1+\alpha^2)}+\frac{J_{\rm AF} \gamma \Delta \sin(\phi_{\rm L}-\phi_{\rm U})}{M(1+\alpha^2)}\mp\frac{K_{\rm h}\gamma \Delta \sin (2\phi_{\rm U,L})}{M(1+\alpha^2)},\label{dq_dt}\\
\dot\phi_{\rm U,L}&=&\pm\frac{u(\alpha-\beta)}{\Delta (1+\alpha^2)}\mp\frac{B_{\rm SO}\gamma }{(1+\alpha^2)}\pm\frac{J_{\rm AF} \alpha\gamma \sin(\phi_{\rm L}-\phi_{\rm U})}{M(1+\alpha^2)}-\frac{K_{\rm h}\alpha\gamma\sin (2\phi_{\rm U,L})}{M(1+\alpha^2)}.\label{phiU_eq}
\end{eqnarray}
The requirement of $q_{\rm L}=q_{\rm U}\equiv q$ (assumption (i) above) leads from the first equation to $\sin (2\phi_{\rm U})=-\sin(2\phi_{\rm L})$, with two solutions as (i) $\phi_{\rm U}=-\phi_{\rm L}$, or (ii) $\phi_{\rm U}=\phi_{\rm L} + (2n+1)\pi/2$ with $n$ being integers. Only solution~(i) can simultaneously fulfill the second equation for $\dot{\phi}_{l}$. After static motion, setting $\dot{\phi}_l=0$, we get from the second equation,
\begin{eqnarray}
0&=&\frac{u(\alpha-\beta)}{\Delta (1+\alpha^2)}+\frac{-\gamma B_{\rm SO} }{(1+\alpha^2)}-\frac{\alpha\gamma (J_{\rm AF}+K_{\rm h})}{M(1+\alpha^2)}\sin (2\phi_{\rm U}),\\
\sin(2\phi_{\rm U})&=&\frac{-M}{\alpha\gamma(J_{\rm AF}+K_{\rm h})}\Big[ \frac{u(\beta-\alpha)}{\Delta(\dot{q}) }+\gamma B_{\rm SO} \Big].~\label{sin2phiU}
\end{eqnarray}
This is Eq.~(1) in the main text.
Note that for self-consistency, in the end we need to plug $\Delta(\dot{q}_l)$ into this equation to check whether $-1\le\sin(2\phi_l)\le 1$, the condition for Walker limit that corresponds to the existence of a constant-$\phi_l$ DW during motion, since this equation comes from setting $\dot{\phi}_l=0$. For now, we first assume that this condition is fulfilled and plug the term $\sin(2\phi_{\rm U})$ into Eq.~(\ref{dq_dt}) to get
\begin{eqnarray}
\dot{q}&=&u\frac{\beta}{\alpha}+\frac{B_{\rm SO}\gamma\Delta(\dot{q})}{\alpha}=u\frac{\beta}{\alpha}+\frac{B_{\rm SO}\gamma a_0}{\alpha}\sqrt{\frac{J_{\rm F}}{2K_{\rm e}}\Big(1-\frac{\dot{q}^2}{v^2_{g}}\Big)}.~\label{DW_velocity}\\
\Rightarrow \dot{q}&=&\frac{2u \alpha \beta K_{\rm e} J_{\rm AF}}{(\hbar \gamma B_{\rm SO})^2  + 2K_{\rm e} J_{\rm AF}\alpha^2}\pm\frac{\gamma B_{\rm SO} \sqrt{J_{\rm AF}\Big[J_{\rm F}(a_0\hbar\gamma B_{\rm SO})^2+2J_{\rm F}J_{\rm AF}K_{\rm e}(\alpha a_0)^2-2K_{\rm e}(u\beta \hbar)^2\Big]}}{(\hbar \gamma B_{\rm SO})^2  + 2K_{\rm e} J_{\rm AF} \alpha^2}.\label{v_solution}
\end{eqnarray}
where $v_{g}=a_0\sqrt{J_{\rm F}J_{\rm AF}}/\hbar$ is the maximal magnon group velocity of the model derived in SI~Sec.~\ref{app}. This is Eq.~(3) in the main text.

In the following we show the necessity of taking $x=q_l$. If we were not taking this substitution, since both sides in Eq.~(\ref{qU_dot_Delta},\ref{dtphiU}) should be independent of $x$, the requirements that the factors of $\text{tanh}(\frac{x-q_{l}}{\Delta})$ should be zero leads to the equation
\begin{eqnarray}
\cos (2\phi_{\rm U})&=&\cos (2\phi_{\rm L})=\frac{-J_{\rm F}a^2_0+\Delta^2(J_{\rm AF}+2K_{\rm e}+K_{\rm h})-J_{\rm AF}\Delta^2\cos(\phi_{\rm L}-\phi_{\rm U})}{K_{\rm h}\Delta^2},~\label{cos1}
\end{eqnarray}
while the requirement of $q_{\rm L}=q_{\rm U}$ (assumption (i) above), when considering $x$-independent parts in above equations, leads to $\sin (2\phi_{\rm U})=-\sin(2\phi_{\rm L})$, with two solutions as (i) $\phi_{\rm U}=-\phi_{\rm L}$, or (ii) $\phi_{\rm U}=\phi_{\rm L} + (2n+1)\pi/2$ ($n=$ integers).
Only solution (i) can satisfy the two equalities in above equation, which can be solved to get
\begin{eqnarray}
\cos (2\phi_{\rm U})&=&\cos (2\phi_{\rm L})=\frac{-J_{\rm F}a^2_0+\Delta^2(J_{\rm AF}+2K_{\rm e}+K_{\rm h})}{(J_{\rm AF}+K_{\rm h})\Delta^2}~\label{cos}.
\end{eqnarray}
This is actually an equation that determines the static DW width $\Delta$. When the current is absent, $u=0$, the DW is in the steady state, thus $\dot{q}_{\rm U}=0$. From Eq.~(\ref{qU_dot_Delta}) and above solution (i), it requires $\phi_{\rm U}=\phi_{\rm L}=0$ which means the DWs are lying on $xz$ plane only. From the above equation with $\phi_{\rm U}=0$ such that $\cos(2\phi_{\rm U})=1$, we get
\begin{eqnarray}
1&=&\cos (2\phi_{\rm L})=\frac{-J_{\rm F}a^2_0+\Delta^2(J_{\rm AF}+2K_{\rm e}+K_{\rm h})}{(J_{\rm AF}+K_{\rm h})\Delta^2}
\Rightarrow\Delta=a_0\sqrt{\frac{J_{\rm F}}{2K_{\rm e}}},
\end{eqnarray}
which is simply the well-known static DW width. This is expectable since we plugged $\theta_{\rm L}=-2\tan^{-1}[\exp\frac{x-q_{\rm L}(t)}{\Delta(\dot{q}_{\rm L})}],\ \theta_{\rm U}=\theta_{\rm L}+\pi$ into the LLGS equation. When there is no current this is simply the saddle-point solution of the Hamiltonian containing only the $J_{\rm F}$ and $K_{\rm e}$ terms, thus the $x$-dependent $\text{tanh}(\frac{x-q_{l}}{\Delta})$ term in Eq.~(\ref{qU_fac}) should vanish by using the form of static DW width. The additional $J_{\rm AF}$ and $K_{\rm h}$ terms do not affect this saddle-point solution, since they favor the DWs to be antiferromagnetically coupled and to lie on $xz$ plane, and the lowest-energy configuration is still described by this $\theta_{l}$.

However, when there is a finite current, $u\neq 0$, we will see Eq.~(\ref{cos}) is not consistent with the terminal solution of $\sin(2\phi_{\rm U})$. Setting $\dot{\phi}_{\rm U}=0$ in Eq.~(\ref{dtphiU}) and from Eq.~(\ref{qU_dot_Delta}), we get (set $B_{\rm SO}=uf_{\rm SO}$ as described in the main text)
\begin{eqnarray}
\sin(2\phi_{\rm U})&=&\frac{-uM}{\alpha\gamma(J_{\rm AF}+K_{\rm h})}\Big[ \frac{\beta-\alpha}{\Delta}+\gamma f_{\rm SO} \Big],\\
\dot{q}_l\equiv v&=&u\Big(\frac{\beta}{\alpha}+\frac{\gamma f_{\rm SO}\Delta}{\alpha}\Big)\Rightarrow u=\frac{\alpha v}{\beta+\gamma f_{\rm SO}\Delta}\Rightarrow \sin(2\phi_{\rm U})=\frac{-vM(\beta-\alpha+\gamma f_{\rm SO}\Delta)}{\gamma \Delta(J_{\rm AF}+K_{\rm h})(\beta+\gamma f_{\rm SO}\Delta)}.~\label{sin2phiU2}
\end{eqnarray}
We consider the Lorentz-contracted DW width as
\begin{eqnarray}
\Delta&=&a_0\sqrt{\frac{J_{\rm F}}{2K_{\rm e}}}\sqrt{1-\frac{v^2}{v^2_g}}\Rightarrow v=v_g\sqrt{1-\frac{2K_{\rm e}\Delta^2}{J_{\rm F}a^2_0}}.~\label{v_delta}
\end{eqnarray}
Using Eq.~(\ref{v_delta}), we can express $\sin(2\phi_{\rm U})$ in Eq.~(\ref{sin2phiU2}) as a function of $\Delta$ only, without the explicit dependence on $v$ and $u$. Now we can check whether the mathematical rule $F(\Delta)\equiv\sin^2(2\phi_{\rm U})+\cos^2(2\phi_{\rm U})-1=0$ is satisfied or not by using Eq.~(\ref{sin2phiU2},\ref{cos}). After some algebra,
\begin{eqnarray}
F(\Delta)&=&\frac{(J_{\rm F}a^2_0 - 2 K_{\rm e} \Delta^2)(A+v^2_g M^2 B)}{(a_0\gamma \Delta^2)^2 J_{\rm F} (J_{\rm AF} + K_{\rm h})^2 (\beta + \gamma f_{\rm SO}\Delta)^2},\\
A&=&a^2_0\gamma^2 J_{\rm F}(\beta + \gamma f_{\rm SO} \Delta)^2 (J_{\rm F}a_0^2 - 2 (J_{\rm AF} + K_{\rm e} + K_{\rm h}) \Delta^2),\ B=\Delta^2 (\beta-\alpha + \gamma f_{\rm SO} \Delta)^2.\nonumber
\end{eqnarray}
The solution of $\Delta$ to fulfill the condition $F(\Delta)=0$ is $\Delta=a_0\sqrt{\frac{J_{\rm F}}{2K_{\rm e}}}$, which is the static DW width. However, this solution contradicts with the Lorentz-contracted width in Eq.~(\ref{v_delta}) which we already used for the derivation of $F(\Delta)$. Therefore, it means we cannot fulfill the condition $\sin^2(2\phi_{\rm U})+\cos^2(2\phi_{\rm U})=1$ using Eq.~(\ref{sin2phiU2}) and Eq.~(\ref{cos}). This means that, $\theta_{\rm L}=-2\tan^{-1}[\exp\frac{x-q_{\rm L}(t)}{\Delta(\dot{q}_{\rm L})}],\ \theta_{\rm U}=\theta_{\rm L}+\pi$ are no longer the the saddle-point solutions when the external current is finite. To circumvent this difficulty, we focus on DW center (as in \cite{Parkin,Mougin}), then Eq.~(\ref{sin2phiU2}) becomes the only approximate solution of $\sin(2\phi_{\rm U})$, then the above Eq.~(\ref{sin2phiU}) and Eq.~(\ref{DW_velocity}) follow.
\subsection{Large current limit of DW velocity}
Dependence on current $u$ of the DW velocity can be seen as (setting $\dot{q}=v$)
\begin{eqnarray}
v&=&\frac{u (c_3+c_4\sqrt{1+c_5 u^2})}{c_1+c_2 u^2},\ c_1=2K_{\rm e} J_{\rm AF}\alpha^2,\ c_2=(\hbar \gamma f_{\rm SO})^2,\ ~\label{q(u)}\\
c_3&=&2 \alpha \beta K_{\rm e} J_{\rm AF},c_4=\alpha\gamma f_{\rm SO}  J_{\rm AF}a_0\sqrt{2 J_{\rm F} K_{\rm e}},\ c_5=\frac{\hbar^2 [J_{\rm F}(\gamma f_{\rm SO} a_0)^2 - 2 K_{\rm e}\beta^2]}{2 J_{\rm AF} J_{\rm F} K_{\rm e} (\alpha a_0)^2},\nonumber
\end{eqnarray}
where we have used $B_{\rm SO}=f_{\rm SO}u$ from density-functional calculation~\cite{Wadley}.
More concisely, we find the DW velocity can be exactly reduced to the form
\begin{eqnarray}
v&=&\frac{u}{a\sqrt{c+du^2}-b}~\label{v_of_u},\ a=\gamma f_{\rm SO}/(FJ_{\rm AF}),\ b=2 K_{\rm e} \alpha \beta/F,\\
c&=&2 J_{\rm AF}^2 J_{\rm F} K_{\rm e} (a_0\alpha)^2,\ d=J_{\rm AF}\hbar^2 F,\ F=J_{\rm F}  (\gamma f_{\rm SO} a_0)^2-2 K_{\rm e} \beta^2.\nonumber
\end{eqnarray}
If we fix $\beta=2\alpha$, the dependence of $v$ on $u$ is illustrated in Supplementary Fig.~\ref{v_current}~(a). At present stage, we ignore the Walker breakdown, and assume this formula is exact for all $u$. Then in the large $u$ limit, we get
\begin{eqnarray}
\lim_{u\rightarrow\infty}v\rightarrow\frac{c_4\sqrt{c_5}}{c_2}&=&\frac{a_0}{\hbar}\sqrt{J_{\rm F}J_{\rm AF}}\sqrt{1-\frac{2K_{\rm e}}{J_{\rm F}}\Big(\frac{\beta}{\gamma f_{\rm SO}a_0}\Big)^2}<v_{g}=\frac{a_0}{\hbar}\sqrt{J_{\rm F}J_{\rm AF}},
\end{eqnarray}
with $\frac{2K_{\rm e}}{J_{\rm F}}\Big(\frac{\beta}{\gamma f_{\rm SO}a_0}\Big)^2\approx0.0004$ using our parameters with $\beta=10\alpha$. Therefore, the DW velocity cannot exceed the maximal magnon velocity as expected, and $\lim_{u\rightarrow\infty}v\rightarrow v_{g}$ only when $\beta=0$, namely the adiabatic limit.
\begin{figure}[tb]
	\centering
	\includegraphics[scale=0.4]{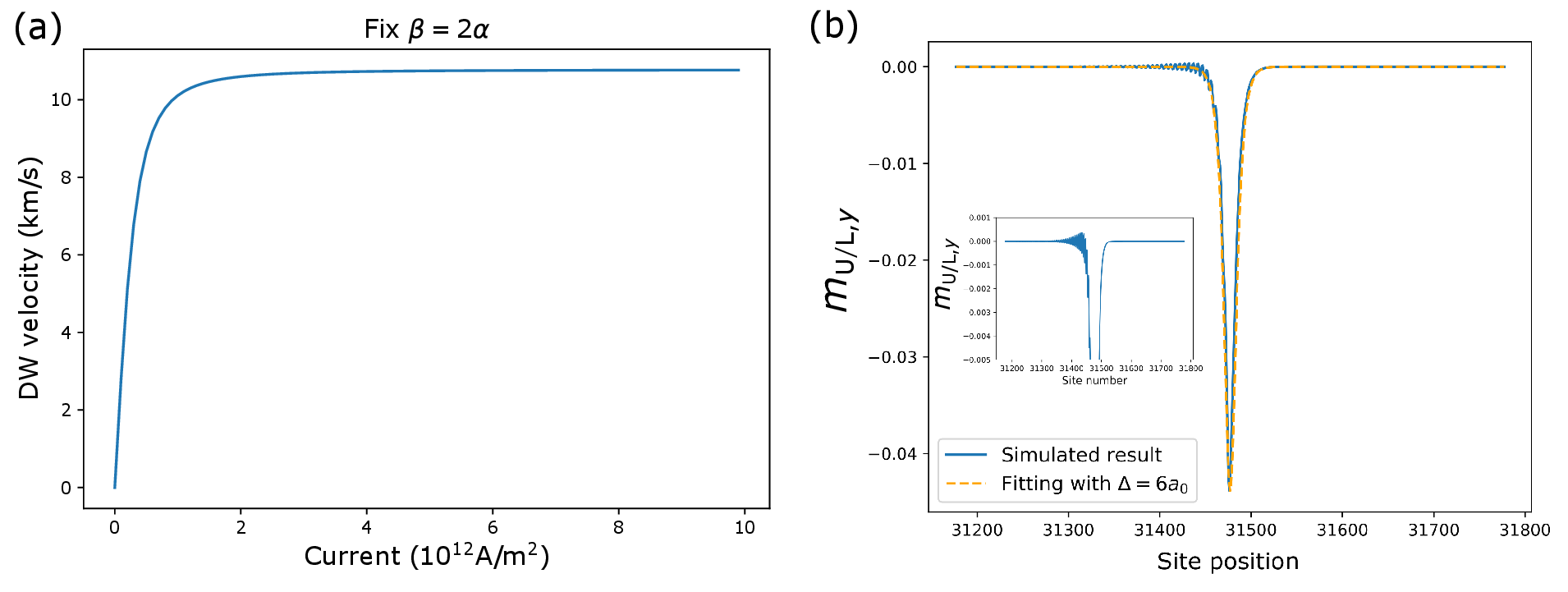}\renewcommand{\figurename}{Supplementary Fig.}\setcounter{figure}{0}
	\caption{(a) Domain Wall (DW) velocity vs current density, with $\beta=2\alpha$. (b) Hard-axis tilted magnetization component $m_{ \text{U/L},y}$ as a function of the site position after applying the current for 1000 ps. The blue curve represents the $m_{ \text{U/L},y}$ for both upper and lower layers (they have nearly the same value), while orange dotted curve shows a fit with the saddle-point DW profile taking $\Delta=6a_0$ and peak magnitude as that of simulated $m_{ \text{U/L},y}$. Inset shows the enlarged view of the spin waves behind the DW.}
	\label{v_current}
\end{figure}
\section{Magnon velocity and DW hard-axis tilt angle}
In addition to the good agreement among the analytical and simulated DW velocities as shown in Fig.~1 in the main text, we find the analytical hard-axis tilt angle (tilted in $y$-direction in Supplementary Fig.~\ref{v_current}~(b)) can also match the simulated one. For example, with $J=1.5\times 10^{12}$~A/m$^2$, in the presence of both SOT and STT, from Eq.~(\ref{sin2phiU}) we find $\sin(2\phi_{\rm U})=-0.16$. The value of $\sin(2\phi_{\rm U})$ corresponds to an angle $\phi_{\rm U,L}\approx \mp3.63^\circ$, inducing a nonzero normalized $m_{\text{U/L},y}=\sin\theta_{\rm U,L}\sin\phi_{\rm U,L}\approx -0.07\text{sech}(\frac{x-vt}{\Delta(v)})$.
This analytical tilted component matches well with simulation as shown in Supplementary Fig.~\ref{v_current}~(b), as (i) simulated $m_{\text{U/L},y}$ has the same negative sign for both upper and lower layers, (ii) the shape of $m_{\text{U/L},y}$ is of a form of hyperbolic secant centered nearly on the DW center, and (iii) the minor discrepancy is that numerically the minimal value of $m_{\text{U/L},y}$ is $-0.044$, about only half of the analytical value $-0.08$, but with the same order of magnitude. This may be attributed to nonzero spin wave emission behind the moving DW as can be seen by the ripples on the left side of the DW profile in Supplementary Fig.~\ref{v_current}~(b).
We have used a saddle-point solution of $\theta_{\rm U,L}$ to fit the numerical DW width during motion, which takes a value of $\Delta_{\rm fit}\approx 6a_0$. From this fitted width and the velocity $v$ measured by the position of peak of $m_{\text{U},x}$, which is $v\approx10.5$ km/s, we can extract the numerical $v_g$ as $v_g=v/\sqrt{1-\Delta^2_{\rm fit}/\Delta_0}\approx 10.83$ km/s, with $\Delta_0\equiv a_0\sqrt{J_{\rm F}/2K_{\rm e}}\approx 25a_0$. This is very close to the analytical one, $v_g=a_0\sqrt{J_{\rm F}J_{\rm AF}}/\hbar=10.77$ km/s. These aspects together with the good prediction of DW velocities support the accuracy of our analytical theory on the DW dynamics.
\section{Calculation of Walker breakdown threshold current}
\begin{figure}[tb]
	\centering
	\includegraphics[scale=0.6]{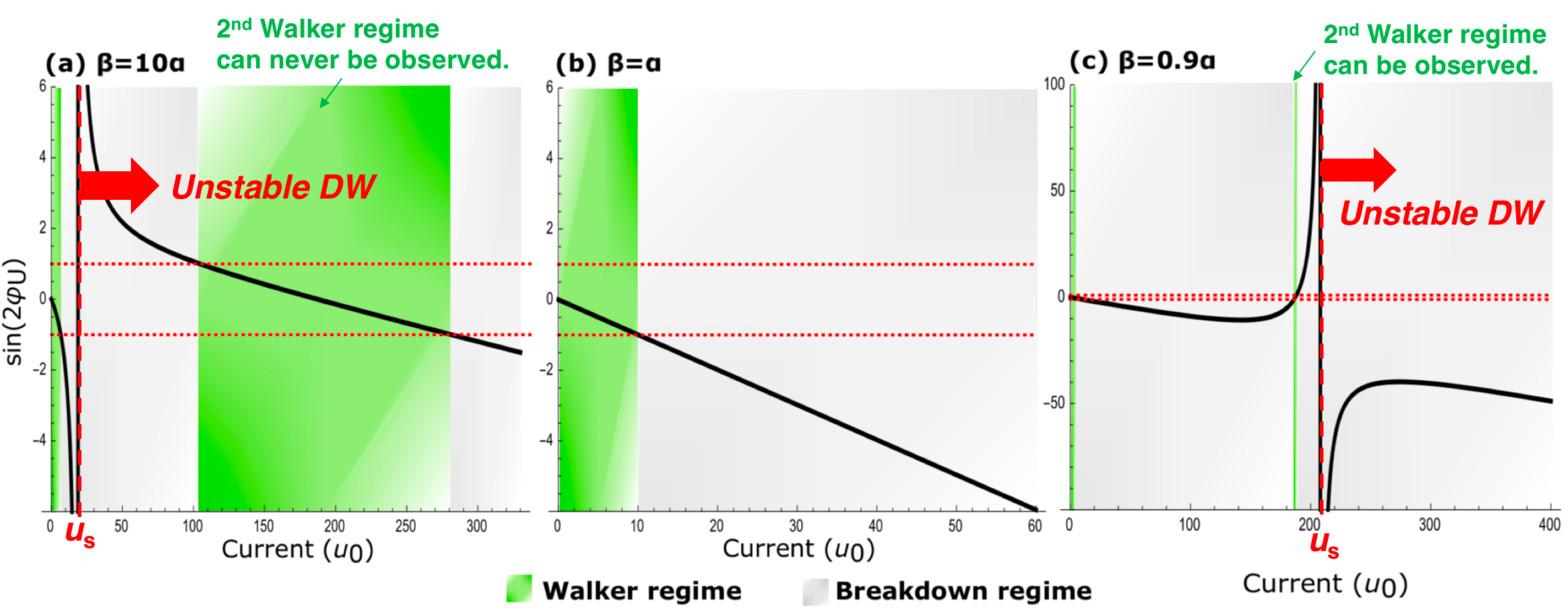}\renewcommand{\figurename}{Supplementary Fig.}
	\caption{The $\sin(2\phi_{\rm U})$ diagram in SI for $\alpha=0.001$ and $J_{\rm AF}$ being the value in Mn$_2$Au. As shown in the main text, with current larger than $u_{\rm s}$, the DW is unstable. Therefore, only in the case with $\beta<\alpha$ can the second Walker regime be observed.}
	\label{sin2phi_}
\end{figure}

To find the threshold current of Walker breakdown, first note from Eq.~(\ref{DW_velocity}) that we can write the DW width as $\Delta(v(u))=\frac{\alpha v(u)-\beta u}{\gamma f_{\rm SO} u}$, then from Eq.~(\ref{sin2phiU},\ref{v_solution}), after some calculations we get
\begin{eqnarray}
\sin(2\phi_{\rm U})&=&-c_0u\Big[1+\frac{(\beta-\alpha)(c_1+c_2 u^2)}{x_1\sqrt{1+c_5 u^2}-x_2 u^2}\Big],\ c_0=\frac{f_{\rm SO} M}{\alpha(J_{\rm AF}+K_{\rm h})},\ x_1=\alpha c_4,\ x_2=\beta c_2,\label{sin1}
\end{eqnarray}
using $c_n$'s defined in Eq.~(\ref{q(u)}). This is one of the major results claimed in the main text. The equation satisfied by the threshold (or termed as critical) current, $\sin(2\phi_{\rm U}(u_{\rm c}))=\pm 1$, has no exact solution of $u_{\rm c}$ since it can be reduced to a sixth-order polynomial equation of $u_{\rm c}$, $\sum^6_{n=1}\gamma_n u^n_{\rm c}=0$ with some tedious coefficients $\gamma_n$ by rearranging terms and taking squares on both sides in Eq.~(\ref{sin1}). 
Supplementary Fig.~\ref{sin2phi_} shows $\sin(2\phi_{\rm U})$ for (a) $\beta>\alpha$, (b) $\beta=\alpha$, and (c) $\beta<\alpha$. Comparing (a) and (c), we find the singular current $u_{\rm s}$ occurs before (behind) the second Walker regime for $\beta>\alpha$ ($\beta<\alpha$). Therefore the second Walker regime can only be observed for the case of $\beta<\alpha$. When $\beta=\alpha$, we obtain $\sin(2\phi_{\rm U})=-c_0u$ from Eq.~(\ref{sin1}), which becomes linearly dependent on $u$, and there is only one threshold current $u_{\rm c}=1/c_0$, above which the system is driven into a unique breakdown regime, similar to the case in ferromagnetic DWs.
\section{Physical mechanism of reentrant Walker regime}
We explain the physical mechanism that drives the system into reentrant Walker regime by considering the dynamics of $\phi\equiv\phi_{\rm U}=-\phi_{\rm L}$ which can be written as
\begin{eqnarray}
&&\dot{\phi}=\frac{1}{1+\alpha^2}\Big[-u\Big(\frac{\beta-\alpha}{\Delta}+\gamma f_{\rm SO}\Big)-\frac{\alpha\gamma(J_{\rm AF}+K_{\rm h})}{M}\sin(2\phi)\Big].\label{phi_eq}
\end{eqnarray}
\textit{Mechanism for the first Walker regime and first Walker breakdown.} Initially without current ($u=0$), $\phi$ is zero to fulfill the DW saddle-point solution. When a finite current, $u>0$, is applied for $\beta>\alpha$, the first torque in the bracket is negative, while initially the second torque is zero. Then the initially zero $\phi$ becomes negative and induces a positive torque from the second term. The two torques in the bracket thus compete, since the AF exchange coupling and $\hat{\bm{y}}$-directional hard axis anisotropy both prefer a zero $\phi$ as the lowest-energy state. When $|\phi|$ becomes large enough, the two torques completely compensate, leading to a constant-$\phi$ motion of the DWs in the Walker regime. 
For a larger $u$, DW velocity increases and its width $\Delta$ decreases, then the first torque decreases. The Walker breakdown occurs when the current is large enough such that the second torque with the maximal magnitude of $\frac{\alpha\gamma(J_{\rm AF}+K_{\rm h})}{M(1+\alpha^2)}$ can no longer compensate the first torque from STT and SOT.

\textit{Mechanism for reentrant Walker regime.} In the breakdown regime, the dynamics of $\phi$ is still governed by Eq.~(\ref{phi_eq}) when ignoring spin wave emissions. Consider the $\beta<\alpha$ case and we write
\begin{eqnarray}
&&\dot{\phi}=\frac{1}{1+\alpha^2}\Big\{-u\Big[\gamma f_{\rm SO}-\frac{(\alpha-\beta)}{\Delta}\Big]-\frac{\alpha\gamma(J_{\rm AF}+K_{\rm h})}{M}\sin(2\phi)\Big\}.\label{phi_eq2}\\
&&\dot{\phi}=0\Rightarrow \sin(2\phi)=\frac{-uM}{\alpha\gamma(J_{\rm AF}+K_{\rm h})}\Big[\gamma f_{\rm SO}-\frac{(\alpha-\beta)}{\Delta}\Big].\label{sin}
\end{eqnarray}
Now the bracket $[\gamma f_{\rm SO}-\frac{(\alpha-\beta)}{\Delta}]$ can be either positive or negative depending on magnitude of $\Delta$. Regardless of its sign, the terminal state with $\dot{\phi}=0$ and Walker breakdown can occur by the same argument in the previous paragraph. Supposed initially without current the bracket is positive. In Fig.~2 in main text, above $u_{\rm c1}$ we have the RHS of Eq.~(\ref*{sin}) being less than $-1$, which means the bracket is positive. 
To explain the reentrant Walker regime, we numerically solved the DW dynamics by RK method (see below), with the time-averaged velocity shown as black dots in Supplementary Fig.~\ref{sin2phi}.

\begin{figure}[tb]
	\centering
	\includegraphics[scale=0.38]{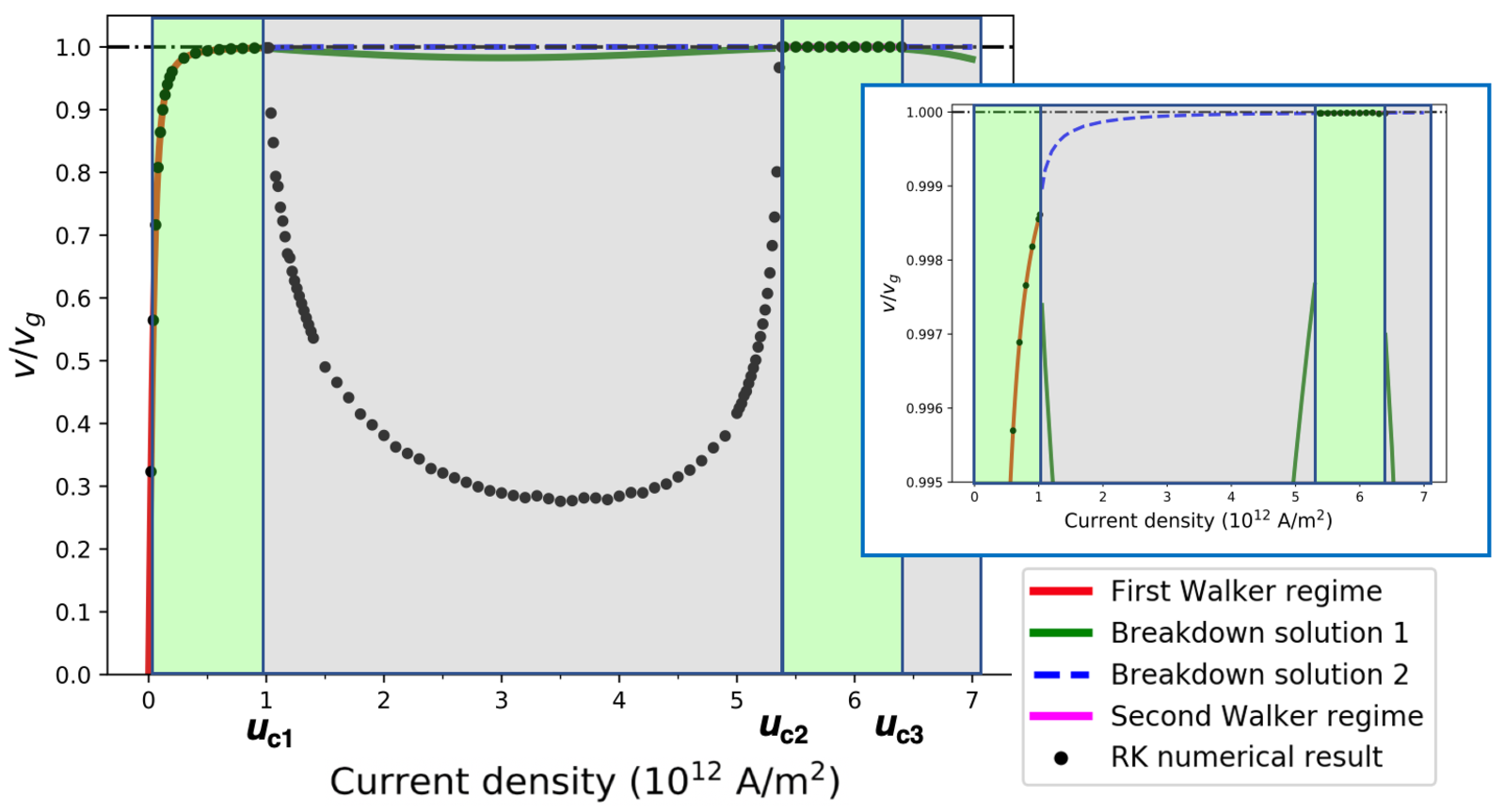}\renewcommand{\figurename}{Supplementary Fig.}
	\caption{DW velocity (averaged velocity) in Walker (breakdown) regimes. Inset shows the enlarged view.}
	\label{sin2phi}
\end{figure}
\begin{figure}[tb]
	\centering
	\includegraphics[scale=0.6]{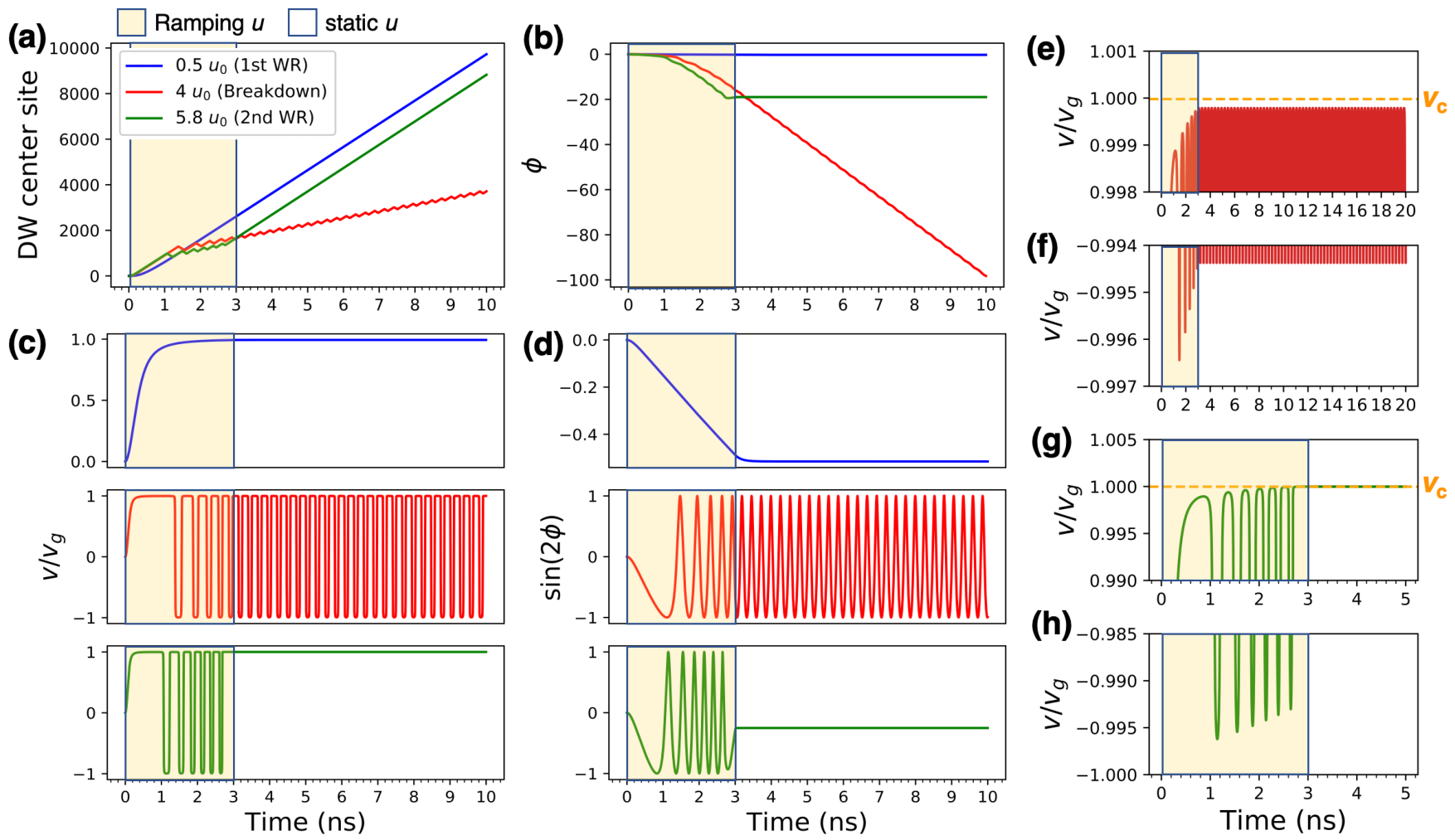}\renewcommand{\figurename}{Supplementary Fig.}
	\caption{Runge-Kutta calculation results for three different current values as $u=0.5u_0, 4 u_0, 5.8u_0$ in the first Walker regime (WR), breakdown regime, and second WR, respectively. (e-h) show the enlarged view of DW velocities.}
	\label{RK1}
\end{figure} 
\begin{figure}[tb]
	\centering
	\includegraphics[scale=0.6]{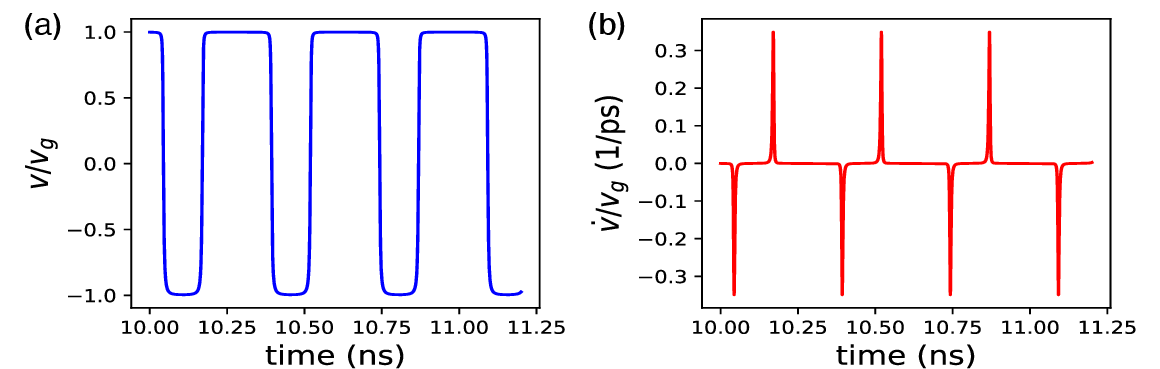}\renewcommand{\figurename}{Supplementary Fig.}
	\caption{Runge-Kutta calculation results for (a) DW velocity $v/v_g$ and (b) acceleration $\dot{v}/v_g$ in breakdown regime with $u=3u_0$.}
	\label{RK2}
\end{figure}
In RK calculation, the current is linearly ramped up from zero to target value in 3~ns. Figure~\ref{RK1} shows the DW position, velocity and tilt angle. In breakdown regime (red curves), both DW velocity ratio $v/v_g$ and tilt angle $\phi$ oscillate in time as expected. 
Interestingly, in the second Walker regime (green curves), although $v/v_g$ and $\phi$ oscillate when the current is ramped through the values in breakdown regime, after current is ramped to a specific value, the oscillation stops, and $v$ and $\phi$ become static again, meaning the Walker regime is retrieved.

How the Walker regime is re-entered can be shown by Supplementary Fig.~\ref{RK1}~(e-h) as enlarged views of velocities. Orange dotted lines show the critical velocity $v_c$ at which the RHS of Eq.~(\ref{sin}) increases from less than $-1$ to exactly $-1$. In Supplementary Fig.~\ref{RK1}~(g), the velocity peak increases when current is ramped, leading to a decreasing $\Delta$, giving the possibility for RHS of Eq.~(\ref{sin}) to increase. Comparing Supplementary Fig.~\ref{RK1}~(e) and (g), we find for current within the second Walker (first breakdown) regime, the velocity can (cannot) reach the critical current $v_c$, leading to the presence (absence) of the second Walker regime.

The origin of this phenomenon is the dependence of STT on domain wall width $\Delta$ in Eq.~(\ref{phi_eq}). The STT is induced by angular momentum conservation when electron spins in the current rotate to align with local magnetizations. The rate of reverse of electron spin is roughly proportional to $u/\Delta$, since $\Delta$ characterizes the length scale of the reverse of local magnetizations in a DW, and $u$ characterizes the polarized electron's velocity. Therefore, there is a $u/\Delta$ dependence in Eq.~(\ref{phi_eq}) that characterizes the strength of STT. Due to the Lorentz contraction of $\Delta$ in antiferromagnets, the competition between STT ($\propto u/\Delta$) and SOT ($\propto u$) in Eq.~(\ref{phi_eq}) contribute to the reentrant Walker regime.

We note the drastic oscillation of $v/v_g$ between $\pm 1$ in Supplementary Fig.~\ref{RK1} is a surprising behavior, since one may expect it as a small oscillation close to $+1$ without crossing into negative values. In Supplementary Fig.~\ref{RK2}~(b-c), the DW acceleration ($\dot{v}/v_g$) shows alternative peaks with a temporal width measured as around 10 ps, which stems from its formula in Eq.~(\ref{f_v}) whose denominator could approach zero. This leads to the oscillating $v/v_g$ between $\pm1$. Realistically, this drastic motion of DW which changes its moving direction in the time scale of 10 ps may excite spin waves that hinder the second Walker regime to be observed experimentally. It is our future scope to include the spin wave excitations in theory.
\section{Runge-Kutta calculation of DW dynamics}
Assuming $\Delta$ is Lorentz contracted relative to the time-varying velocity, we can numerically solve the dynamics of DW center $q(t)$ and tilt angel $\phi(t)$ driven by a time-dependent current $u(t)$ in both Walker and breakdown regimes. Time derivatives of $\phi, q, v$ are
\begin{eqnarray}
\dot{\phi}&=&\frac{1}{1+\alpha^2}\Big\{-u\Big[\gamma f_{\rm SO}-\frac{(\alpha-\beta)}{\Delta}\Big]-\frac{\alpha\gamma(J_{\rm AF}+K_{\rm h})}{M}\sin(2\phi)\Big\},\label{f_phi}\\
\dot{q}=v&=& \frac{1}{1+\alpha^2}\Big\{u(1+\alpha\beta)+\Big(u\alpha\gamma f_{\rm SO}-\frac{\gamma(J_{\rm AF}+K_{\rm h})}{M}\sin(2\phi)\Big)\Delta\Big\},\label{f_q}\\
\dot{v}&=&\frac{1}{1+\alpha^2}\Big\{\dot{u}(1+\alpha\beta)+\alpha\gamma f_{\rm SO}\Big(\dot{u}\Delta+u\dot{\Delta}\Big)-\frac{\gamma(J_{\rm AF}+K_{\rm h})}{M}\Big(2\cos(2\phi)\dot{\phi}\Delta+\dot{\Delta}\sin(2\phi)\Big)\Big\}\label{dot_v}.
\end{eqnarray}
By using $\Delta=\Delta_0\sqrt{1-(v/v_g)^2}$ with $\Delta_0=a_0\sqrt{J_{\rm F}/2K_{\rm e}}$, we get $\dot{\Delta}=-\Delta_0 v\dot{v}/(v_g\sqrt{v^2_g-v^2})$, so the RHS of Eq.~(\ref{dot_v}) for $\dot{v}$ also depends on $\dot{v}$ itself via $\dot{\Delta}$. Rearranging all $\dot{v}$-dependent terms on one side, we obtain the DW acceleration
\begin{eqnarray}
\dot{v}&=&\frac{\dot{u} \Big(A+B\sqrt{1-(v/v_g)^2}\Big)-2C\cos(2\phi)\sqrt{1-(v/v_g)^2} \dot{\phi}}{1+v\Big(Bu-C\sin(2\phi)\Big)\Big/\Big(v^2_g\sqrt{1-(v/v_g)^2}\Big)},\label{f_v}\\
A&=&\frac{1+\alpha\beta}{1+\alpha^2},\ B=\frac{\alpha\gamma f_{\rm SO}\Delta_0}{1+\alpha^2},\ C=\frac{\gamma(J_{\rm AF}+K_{\rm h})\Delta_0}{(1+\alpha^2)M}.
\end{eqnarray}
The three coupled differential equations Eq.~(\ref{f_phi},\ref{f_q},\ref{f_v}) are solved by the fourth-order RK method with integration time step as 0.01~ps. The averaged velocities in Supplementary Fig.~\ref{RK1}~(b) are calculated in the time range from 10 to 40~ns in which the DW motion is already stable.

\subsection{Approximate analytical calculation of DW velocity in breakdown regime}
We can estimate the time-averaged DW velocity $v_{\rm avg}$ by assuming that $\Delta(v(t))$ has reached a terminal value and it depends roughly on the averaged velocity as $\Delta(v(t))\approx \Delta (v_{\rm avg})$ via Lorentz contraction. By solving the differential equation $\dot{\phi}_{\rm U}=A\sin(2\phi_{\rm U})+B$ with constants $A=\frac{-\alpha\gamma (J_{\rm AF}+K_{\rm h})}{M(1+\alpha^2)}$ and $B=\frac{-u}{1+\alpha^2}(\frac{\beta-\alpha}{\Delta}+\gamma f_{\rm SO})$, we obtain the solution $\phi_{\rm U}(t)=\arctan[\frac{-A}{B}+\frac{\sqrt{B^2-A^2}}{B}\tan[\sqrt{B^2-A^2}(t+c)]]$ with an integration constant $c$. Real solutions require $B^2\ge A^2$, which leads to a criterion $u\ge u_{\rm th}= \frac{\alpha \gamma(J_{\rm AF} + K_{\rm h})\Delta(u_{\rm th})}{M [\gamma f_{\rm SO} \Delta(u_{\rm th}) + (\beta-\alpha)]}$. This equation is exactly the same as that for $u_{\rm c1}$, i.e., $\sin(2\phi_{\rm U}(u_{\rm c1}))=-1$, thus we have $u_{\rm th}=u_{\rm c1}$, which allows us to draw $v_{\rm avg}(u)$ in the first breakdown regime starting almost from $u_{\rm c1}$. By averaging over the period $\pi/\sqrt{B^2-A^2}$, we obtain $\langle\sin(2\phi_U)\rangle =(-B-\sqrt{B^2-A^2})/A$~\cite{Yang}. Using this formula, we can obtain
\begin{eqnarray}
v_{\rm avg}&=&\frac{u(1+\alpha\beta)}{1+\alpha^2}+\Big[\frac{u\gamma f_{\rm SO}\alpha}{(1+\alpha^2)}+\frac{-(J_{\rm AF}+K_{\rm h})\gamma\langle \sin(2\phi_{\rm U})\rangle}{M(1+\alpha^2)}\Big]\Delta(v_{\rm avg})\\
&=&u\frac{\beta}{\alpha}+u\frac{\gamma f_{\rm SO}}{\alpha}\Delta(v_{\rm avg})-\frac{\sqrt{\Big(u M [\gamma f_{\rm SO} \Delta(v_{\rm avg})+ (\beta - \alpha) ]\Big)^2-\Big(\alpha \gamma (J_{\rm AF} + K_{\rm h})\Delta(v_{\rm avg})\Big)^2}}{M\alpha(1+\alpha^2)}\nonumber.
\end{eqnarray}
By rearranging terms, $v_{\rm avg}$ satisfies an equation $\sum^4_{n=0}d_n v^n=0$ with some coefficients $d_n$. We numerically find the four solutions of $v_{\rm avg}$. Two of them are negative or complex and thus unphysical (since current is along $+x$ direction), whereas the other two are real and positive and plotted by solid (green) and dotted (blue) curves in Supplementary Fig.~\ref{sin2phi} as a comparison with RK result. We find the analytical approximation (green curve) is only a very crude upper bound of the more accurate RK result, while the other solution (blue curve) is larger than the velocity before breakdown thus should be excluded by the Lagrangian argument in the main text.
\section{Comparison of torques}
For $\phi\equiv\phi_{\rm U}=-\phi_{\rm L}$, we write the magnetization as [define $()\equiv (\frac{x-q}{\Delta})$]
\begin{eqnarray}
&&\bm{M}_{\rm U/L}=M\Big[\text{sech}()\Big(\pm\cos\phi ~\hat{\bm{x}}+\sin\phi~ \hat{\bm{y}}\Big)\pm \text{tanh}()\hat{\bm{z}}\Big].
\end{eqnarray}
After taking a curl with $\bm{M}$ for the LLGS equation; namely, multiplying $\bm{M}\times$ on both sides of it, the STT, SOT, AF exchange torque (AF-T), and hard-axis anisotropy torque (Kh-T) read
\begin{eqnarray}
\text{STT}&=&\frac{-1}{1+\alpha^2}\Big[\frac{(\alpha-\beta)}{M}\bm{M}_{\rm U/L}\times (\bm{u}\cdot\bm{\nabla})\bm{M}_{\rm U/L} +(1+\alpha\beta)(\bm{u}\cdot\bm{\nabla})\bm{M}_{\rm U/L}\Big],~\label{STT1}\\
\text{SOT}&=&\frac{-\gamma}{1+\alpha^2}\Big[\bm{M}_{\rm U/L}\times \bm{B}_{\rm SO, U/L} +\frac{\alpha}{M} \bm{M}_{\rm U/L}\times(\bm{M}_{\rm U/L}\times \bm{B}_{\rm SO, U/L})\Big],~\label{SOT}\\
\text{AF-T}&=&\frac{J_{\rm AF}}{1+\alpha^2}\Big[\bm{M}_{\rm U/L}\times \bm{M}_{\rm L/U} +\frac{\alpha}{M} \bm{M}_{\rm U/L}\times(\bm{M}_{\rm U/L}\times \bm{M}_{\rm L/U})\Big],~\label{AFT}\\
\text{Kh-T}&=&\frac{2K_{\rm h}}{1+\alpha^2}\Big[\bm{M}_{\rm U/L}\times M_{\rm U/L,y}\hat{\bm{y}} +\frac{\alpha}{M} \bm{M}_{\rm U/L}\times(\bm{M}_{\rm U/L}\times M_{\rm U/L,y}\hat{\bm{y}})\Big].
\end{eqnarray}
Taking $\bm{u}=u\hat{\bm{x}}$, the STT plus SOT can be shown as a summation of following three torques,
\begin{eqnarray}
&&\bm{\tau}_1=\frac{u}{1+\alpha^2}\Big[\frac{(\beta-\alpha)}{\Delta}+\gamma f_{\rm SO}\Big] \bm{M}_{\rm U/L}\times(\pm\hat{\bm{z}}),\label{1st}\label{tau1}\\
&&\bm{\tau}_2=\frac{uM(1+\alpha\beta+\alpha\gamma f_{\rm SO}\Delta)}{(1+\alpha^2)\Delta}\text{sech}()\text{tanh}()\Big(\pm\cos\phi~\hat{\bm{x}}+\sin\phi~ \hat{\bm{y}}\Big),\label{tau2}\\
&&\bm{\tau}_3=\frac{uM(1+\alpha\beta+\alpha\gamma f_{\rm SO}\Delta)}{(1+\alpha^2)\Delta}\text{sech}^2()(\mp\hat{\bm{z}}),\label{tau3}\\
&&\bm{\tau}_2+\bm{\tau}_3=\frac{u(1+\alpha\beta+\alpha\gamma f_{\rm SO}\Delta)}{(1+\alpha^2)\Delta}\bm{M}_{\rm U/L}\times \bm{B}_{\rm eff,U/L},\ \bm{B}_{\rm eff,U/L}=\text{sech}()\Big(\pm\sin\phi~\hat{\bm{x}}-\cos\phi~\hat{\bm{y}}\Big).
\end{eqnarray}
As shown in Supplementary Fig.~\ref*{torques}, supposed $\beta>\alpha$ and $u>0$, then $\bm{\tau}_1$ induces a rotation of $\phi$ towards $-\hat{\bm{y}}$ direction for both layers, while $\bm{\tau}_2$ and  $\bm{\tau}_3$ induce DW movement toward $+\hat{\bm{x}}$ direction for both layers. The AF-T and Kh-T can be summed and separated as following two torques
\begin{eqnarray}
\bm{\tau}_{\rm AF+K_h,1}&=&\frac{(J_{\rm AF}+K_{\rm h})\alpha M}{1+\alpha^2}[\text{sech}^2()\sin(2\phi)\bm{M}_{\rm U/L}\times(\pm\hat{\bm{z}})],\label{AFT1}\\
\bm{\tau}_{\rm AF+K_h,2}&=&\frac{2 (J_{\rm AF}+K_{\rm h}) M^2}{1+\alpha^2}\text{sech}()\text{tanh}()\sin\phi\Big(\mp\hat{\bm{x}}-\alpha~\text{tanh}()\hat{\bm{y}}\Big)\\
&+&\frac{2 (J_{\rm AF}+K_{\rm h}) M^2}{1+\alpha^2}\text{sech}^2()\sin\phi\Big(\cos\phi+\alpha\sin\phi~\text{tanh}()\Big)(\pm\hat{\bm{z}}).
\end{eqnarray}
We note $\bm{\tau}_{\rm AF+K_h,1}$ induces a rotation of $\phi$ to $+\hat{\bm{y}}$ direction when $\phi<0$. It competes with $\bm{\tau}_1$ in Eq.~(\ref{tau1}). These two torques give the expression for $\dot{\phi}$ in Eq.~(\ref{phi_eq2}). Meanwhile, $\bm{\tau}_{\rm AF+K_h,2}$ does not contribute to the rotation of $\phi$ for the central magnetic site located at $x=q(t)$. In our approximation of focusing on the DW center, we do not consider it here for the dynamics of $\phi$.
\begin{figure}[tb]
	\centering
	\includegraphics[scale=0.3]{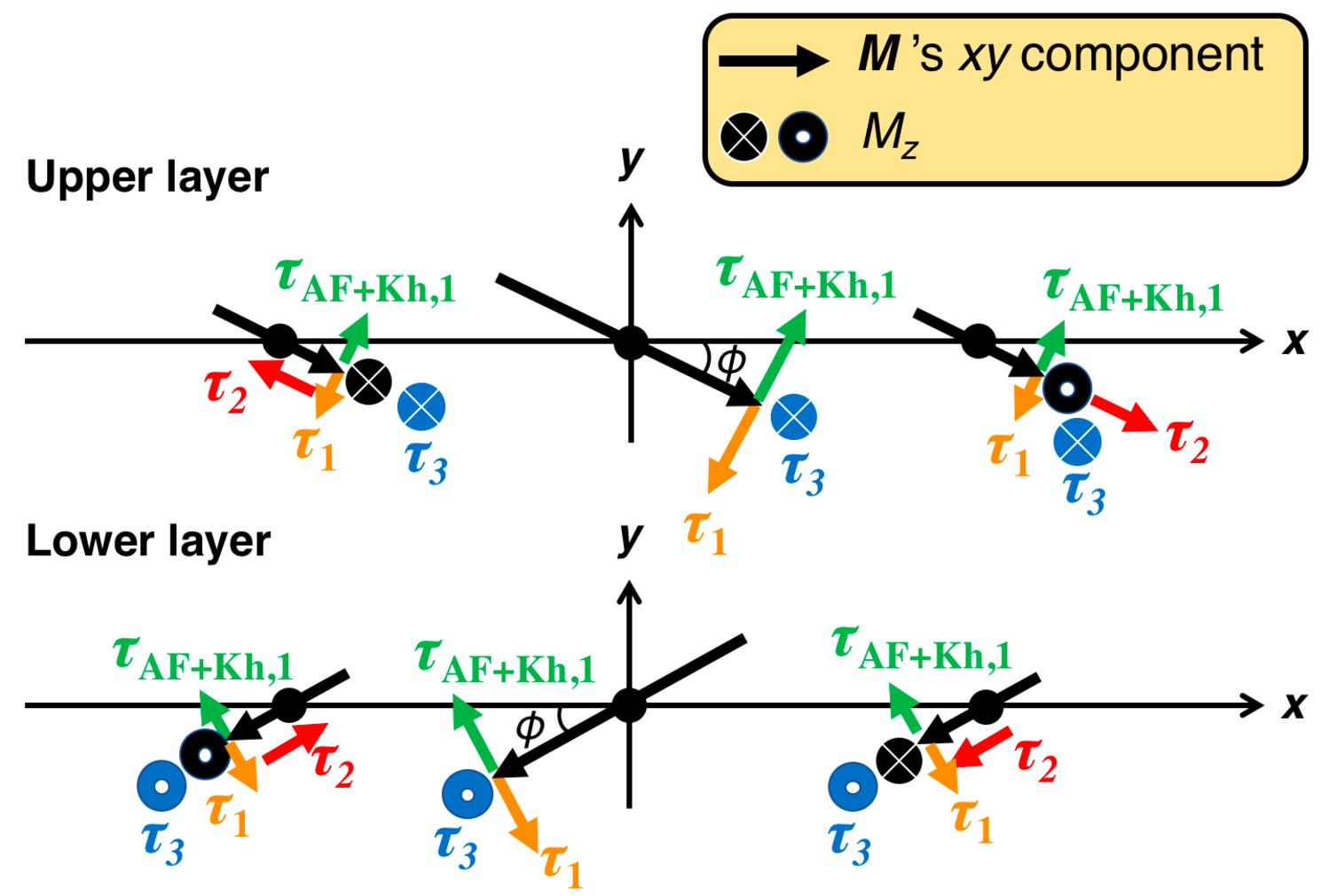}\renewcommand{\figurename}{Supplementary Fig.}
	\caption{Schematics of DW profile and torques with the center located at $x=0$. The black dots and arrows indicate the magnetic moments and their in-plane components, respectively, while colored arrows show the torques.}
	\label{torques}
\end{figure}

To gain some information about the efficiency of STT and SOT, we can estimate their magnitudes shown in Eq.~(\ref{tau1}-\ref{tau3}). The ratio of SOT over STT in their contributions in $\bm{\tau}_1$ is $\gamma f_{\rm SO}\Delta/(\beta-\alpha)\approx 14.7$ using $\alpha=0.001, \beta=10\alpha$, current density as $1.3\times 10^{12}$ A/m$^2$, $B_{\rm SO}=26$ mT, and $\Delta\approx 6.5 a_0$. Therefore, SOT is much efficient to tilt the magnetizations into hard-axis direction and to bring the AF exchange torque between the two layers into effect. 
On the other hand, the ratio between effective fields induced by SOT to that by STT in $\bm{\tau}_2+\bm{\tau}_3$ is $\alpha\gamma f_{\rm SO}\Delta/(1+\alpha\beta)\approx 1.3\times 10^{-4}$, so $\bm{B}_{\rm SOT}$ is much inefficient to push DW compared to $\bm{B}_{\rm STT}$. The major effect by SOT is to efficiently induce hard-axis tilt to generate the large AF exchange torque to drive DW motion, as discussed in the main text.
\begin{figure*}[tb]
	\centering
	\includegraphics[scale=0.45]{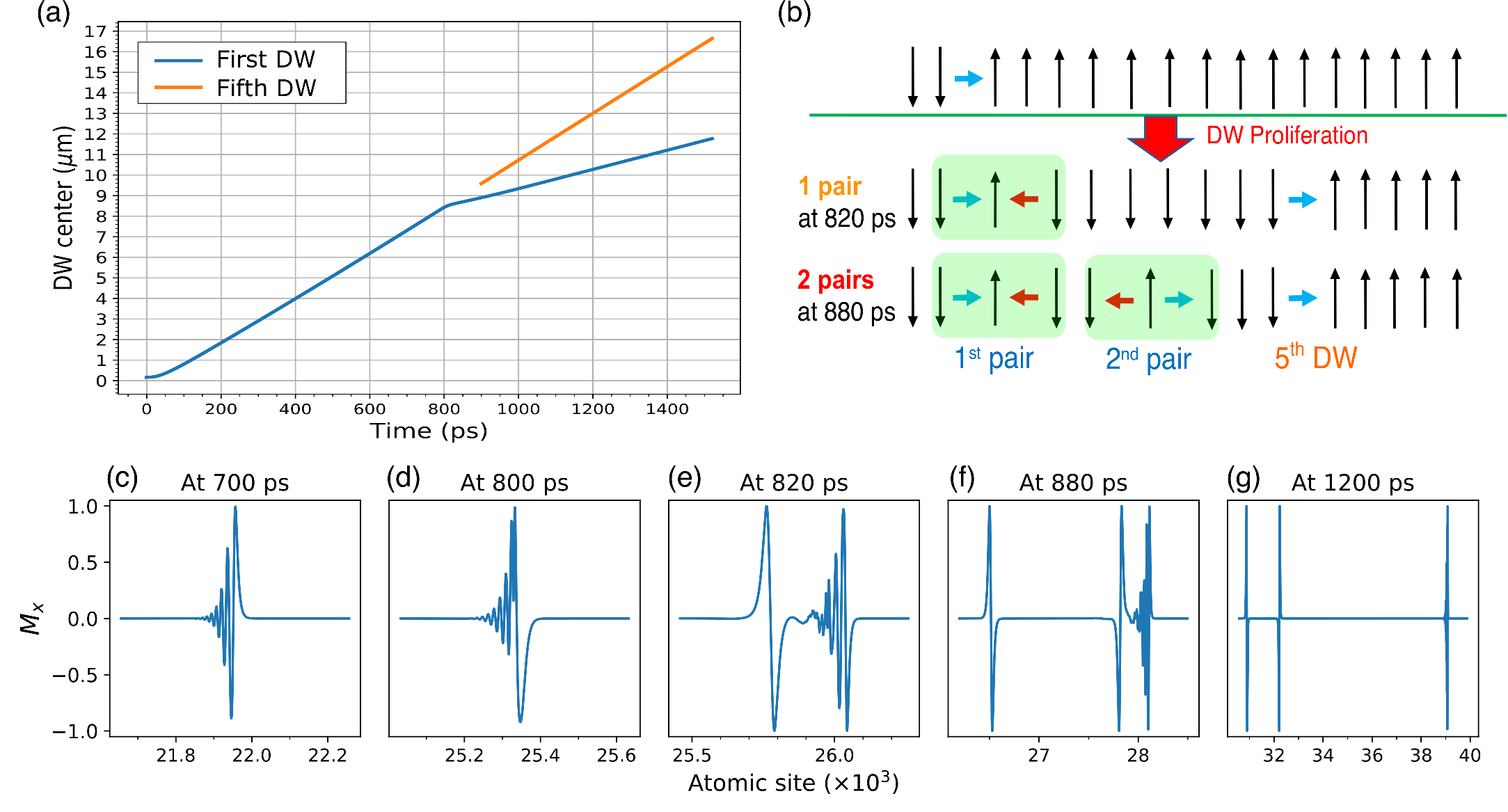}\renewcommand{\figurename}{Supplementary Fig.}
	\caption{Micromagnetic simulation of Domain Wall (DW) motion driven by a current $u=8.125u_0>u_{\rm c1}\approx 6.825u_0$ for $\beta=10\alpha$ (see Supplementary Fig.~\ref{sin2phi_}~(a)) with the current ramp-up time being 1000 ps. (a) DW center as a function of time. (b) Schematics of the magnetization configuration at two temporal instants showing the process of DW proliferations. (c)-(g) Snapshots of the $x$-component of magnetizations, $M_x$, in the upper layer taken at several temporal instants.}
	\label{Fig_WB_motion}
\end{figure*}
\section{Micromagnetic simulation to attempt the breakdown regime}
To test the averaged DW velocity in the breakdown regime, we apply a current larger than $u_{\rm c1}$ to drive the DW motion. We fix $\beta=10\alpha$, since as can be seen by comparing Supplementary Fig.~\ref{Fig_WB} and Fig.~2 in the main text, $u_{\rm c1}$ is smaller in the larger $\beta$ case. Specifically, for $\beta=10\alpha$ we find $u_{\rm c1}\approx 6.825u_0$, and we apply a current with magnitude $u=8.125u_0$. Supplementary Fig.~\ref{Fig_WB_motion} shows the simulated results. In (a), we show the DW center as a function of time. To suppress the spin wave emission or other transient effects due to the high current as much as possible, we use a larger linear ramp-up time as 1000 ps for the current. 

Even in this setup, in Supplementary Fig.~\ref{Fig_WB_motion}~(c) of the $M_x$ snapshot we still find spin waves being emitted behind the moving DW. At 800 ps, the magnitude of applied current is $8.125u_0\times800/1000= 6.5u_0$ which is close to $u_{\rm c1}$. At this instant (d) shows a precursor of negative $M_x$ components in front of the DW, which becomes a new proliferated DW-pair at later times. Approximately after this instant, we find the system starts to proliferate four new DWs (schematically shown in (b)), with two DW bound pairs and a fifth DW moving faster than those two pairs. In (a) the orange curve shows the position of the fifth DW which moves in a velocity of $\approx 1.05v_g$. This Lorentz invariance breaking regime has already been found in~\cite{Ruben2020} and it can be explained by linear momentum transfer among the new born DW-pairs. The two bound DW pairs have nearly the same velocity as about $0.43 v_g$ after 1000 ps, which is much smaller than the  time-averaged velocity by our analytical calculation which is as high as $99.86v_g$. This mismatch must come from our assumption of the ideal system without DW proliferations and spin wave emissions. Therefore, to verify our theory it is tempting to search for more suitable model parameters which can suppress these phenomena, or to generalize our simple results by taking them into account in future studies.
\section{Derivation of the magnon velocity}~\label{app}
In this section, we derive the maximal magnon group velocity, $v_g=a_0\sqrt{J_{\rm F}J_{\rm AF}}/\hbar$, as used in the main text. We define the \textit{unit-length} magnetizations $\bm{M}_{j,k}$ as located at positions $(x,y)=(ja_0,ka_0)$, with $j,k$ being integers. Due to the AF exchange along $\hat{\bm{y}}$, we define the two magnetic sublattices $A$ and $B$ through $\bm{M}_{j,2n}=\bm{M}_{A,j}$ and $\bm{M}_{j,2n+1}=\bm{M}_{B,j}$ with $n$ being integers. The unit-length averaged and staggered magnetizations are, respectively, $\bm{m}_{j,2n}\equiv (\bm{M}_{j,2n}+\bm{M}_{j,2n+1})/2$ and $\bm{l}_{j,2n}\equiv(\bm{M}_{j,2n}-\bm{M}_{j,2n+1})/2$, defined per two lattice constants along $\hat{\bm{y}}$. The AF exchange energy can be written as (set $a_0=1$ in subindices for brevity)
\begin{eqnarray}
E_{\rm AF}&=&J_{\rm AF}\sum_{i_x=j\atop i_y=n}\bm{M}_{i}\cdot\bm{M}_{i+\hat{y}}=J_{\rm AF}\sum_{i_x=j\atop i_y=2n}\Big[\bm{M}_{i}\cdot\bm{M}_{i+\hat{y}}+\bm{M}_{i+\hat{y}}\cdot\bm{M}_{i+2\hat{y}}\Big]\\
&=&J_{\rm AF}\sum_{i_x=j\atop i_y=2n}\Big[(\bm{m}_{i}+\bm{l}_{i})\cdot(\bm{m}_{i}-\bm{l}_{i})+(\bm{m}_{i}-\bm{l}_{i})\cdot(\bm{m}_{i+2\hat{y}}+\bm{l}_{i+2\hat{y}})\Big]=2J_{\rm AF}\sum_{i_x=j\atop i_y=2n}\Big[\bm{m}_i^2-\bm{l}_i^2\Big]=4J_{\rm AF}\sum_{i_x=j\atop i_y=2n}\bm{m}_i^2+\text{const.}\nonumber
\end{eqnarray}
where we've used $\bm{m}_{i+2n\hat{y}}=\bm{m}_i,\ \bm{l}_{i+2n\hat{y}}=\bm{l}_i$ for $n\in\mathcal{N}$. In the second step we intentionally restricted the $i_y$ index from any integers to the even-integer subset, since $\bm{m}_i$ and $\bm{l}_i$ fields are defined per two lattice constants in $\hat{\bm{y}}$ direction. In this way the unit cell is doubly expanded in $\hat{\bm{y}}$ direction. The last constant term will be dropped. The ferromagnetic exchange energy reads
\begin{eqnarray}
E_{\rm F}&=&-J_{\rm F}\sum_{i_x=j\atop i_y=n}\bm{M}_{i}\cdot\bm{M}_{i+\hat{x}}=-J_{\rm F}\sum_{i_x=j\atop i_y=2n}\Big[\bm{M}_{i}\cdot\bm{M}_{i+\hat{x}}+\bm{M}_{i+\hat{y}}\cdot\bm{M}_{i+\hat{y}+\hat{x}}\Big]\nonumber\\
&=&-J_{\rm F}\sum_{i_x=j\atop i_y=2n}\Big[(\bm{m}_{i}+\bm{l}_{i})\cdot(\bm{m}_{i+x}+\bm{l}_{i+x})+(\bm{m}_{i}-\bm{l}_{i})\cdot(\bm{m}_{i+\hat{x}}-\bm{l}_{i+\hat{x}})\Big]\nonumber\\
&\approx&-J_{\rm F}\sum_{i_x=j\atop i_y=2n}\Big[2(\bm{m}_i^2+\bm{l}_i^2)+a^2_0\bm{l}_i\cdot (\partial^2_x \bm{l}_i)\Big]\rightarrow J_{\rm F}a^2_0\sum_{i_x=j\atop i_y=2n}(\partial_x\bm{l}_i)^2~\label{J3term}
\end{eqnarray}
where we have Taylor expanded the fields up to second-order of lattice constant $a_0$, and ignored the term $a^2_0\bm{m}_i\cdot(\partial^2_x \bm{m}_i)$ in the exchange limit in which we assume $|\bm{m}_i|\ll |\bm{l}_i|$, and in the last step we have done a partial integration and dropped the constant proportional to $\bm{m}_i^2+\bm{l}_i^2=1$. Now the total exchange energy is
\begin{eqnarray}
E_{\rm F}+E_{\rm AF}&=&\sum_{i_x=j\atop i_y=2n}w_{\text{ex},i}=\sum_{i_x=j\atop i_z=2n}\Big[4J_{\rm AF}\bm{m}^2_i+J_{\rm F}a^2_0(\partial_x\bm{l}_i)^2\Big].~\label{exchange}
\end{eqnarray}
The LLG equations for $\bm{m}_i, \bm{l}_i$ in the exchange limit have the form
\begin{eqnarray}
\dot{\bm{m}}_i&=&\Big(\frac{1}{2}\gamma \bm{B}^{\rm eff}_{l,i}-\alpha \dot{\bm{l}}_i\Big)\times \bm{l}_i\label{LLG_m},\ \dot{\bm{l}}_i= \frac{1}{2}\gamma\bm{B}^{\rm eff}_{m,i}\times \bm{l}_i.
\end{eqnarray}
The derivation of these LLG equations is shown in the bottom of SI. Now we plug the energy summand $w_{\text{ex},i}$ in Eq.~(\ref{exchange}) together with the anisotropy and SO-field terms into above equations to get, from $\bm{B}^{\rm eff}_{l,m}=\frac{-1}{2\gamma\hbar}\partial[w_{\text{ex},i}+(K \text{ terms})+(B_{\rm SO}\text{ term} )]/\partial (\bm{l},\bm{m})$ (denominator $2\gamma\hbar$ comes from $\mu_0\mu_{\rm s}=2\gamma\hbar$ with $\mu_{\rm s}=4\mu_{\rm B}$ for Mn$_2$Au), ignoring site index $i$,
\begin{eqnarray}
\dot{\bm{l}}&=&\frac{1}{2}\gamma\bm{B}^{\rm eff}_m\times \bm{l}=\frac{-2J_{\rm AF}}{\hbar}\bm{m}\times \bm{l}\Rightarrow\bm{m}=\frac{\hbar}{2J_{\rm AF}}\dot{\bm{l}}\times\bm{l},\\
\dot{\bm{m}}&=&\Big(\frac{1}{2}\gamma\bm{B}^{\rm eff}_l-\alpha \dot{\bm{l}}\Big)\times\bm{l}=\frac{\hbar}{2J_{\rm AF}}\ddot{\bm{l}}\times\bm{l}=\Big\{\frac{J_{\rm F}a^2_0}{2\hbar}(\partial^2_x\bm{l})+(K\text{ terms})+(B_{\rm SO} \text{ term})-\alpha\dot{\bm{l}}\Big\}\times \bm{l},\nonumber\\
&\Rightarrow& \bm{l}\times\Big\{(\partial^2_x\bm{l})+\frac{-\hbar^2}{J_{\rm F}J_{\rm AF}a^2_0}(\partial^2_t\bm{l})+(K\text{ terms})+(B_{\rm SO} \text{ term})-\frac{\alpha\hbar}{a^2_0J_{\rm F}}\dot{\bm{l}}\Big\}=0.
\end{eqnarray}
Apart from the $B_{\rm SO}$ and Gilbert damping terms, this equation of motion for $\bm{l}$ is Lorentz invariant. Note that when the DW is moving at steady velocity, both the damping term and the Zeeman term (from $B_\text{SO}$) compensate each other resulting in a prototypical sine-Gordon type of equation. The propagating waves of $\bm{l}$ can be excited, and the first two terms defines the ``speed of light" in this system as
\begin{eqnarray}
v_{g}&=&\frac{a_0\sqrt{J_{\rm F}J_{\rm AF}}}{\hbar}\approx 10.78 \text{ km/s},
\end{eqnarray}
which is the  magnon velocity that matches very well with the simulated result observed by fitting the DW width with its saddle-point-solution form as shown in SI.~Sec.~IV.\\

\textbf{Derivation of the LLG equation for $\bm{m}$ \& $\bm{l}$}

The only energy term that needs a careful treatment is $E_{\rm AF}$, which couples magnetizations at different $y$ points,
\begin{eqnarray}
E_{\rm AF}&=&J_{\rm AF}\sum_{i_x=j\atop i_y=2n}\Big[\bm{M}_{i}\cdot\bm{M}_{i+\hat{y}}+\bm{M}_{i+\hat{y}}\cdot\bm{M}_{i+2\hat{y}}\Big]=J_{\rm AF}\sum_{i_x=j\atop i_y=2n}\Big[(\bm{m}_{i}+\bm{l}_{i})\cdot(\bm{m}_{i}-\bm{l}_{i})+(\bm{m}_{i}-\bm{l}_{i})\cdot(\bm{m}_{i+2\hat{y}}+\bm{l}_{i+2\hat{y}})\Big].\nonumber
\end{eqnarray}
From the derivative of the first expression we get
\begin{eqnarray}
\bm{B}^{\rm eff}_{j,2n}&=&\frac{-\partial E_{\rm AF}}{\partial \bm{M}_{j,2n}}=-J_{\rm AF}\Big(\bm{M}_{j,2n+1}+\bm{M}_{j,2n-1}\Big)\label{H2n},\ \bm{B}^{\rm eff}_{j,2n+1}=\frac{-\partial E_{\rm AF}}{\partial \bm{M}_{j,2n+1}}=-J_{\rm AF}\Big(\bm{M}_{j,2n}+\bm{M}_{j,2n+2}\Big),\label{H2np1}
\end{eqnarray}
while from the derivative of the second expression we get
\begin{eqnarray}
\bm{B}^{\rm eff}_{m,i}&=&\frac{-\partial E_{\rm AF}}{\partial \bm{m}_{i}}=-J_{\rm AF}\Big(2\bm{m}_i+\bm{m}_{i-2\hat{y}}+\bm{m}_{i+2\hat{y}}-\bm{l}_{i-2\hat{y}}+\bm{l}_{i+2\hat{y}}\Big)\label{Hm}\\
\bm{B}^{\rm eff}_{l,i}&=&\frac{-\partial E_{\rm AF}}{\partial \bm{l}_{i}}=-J_{\rm AF}\Big(\bm{m}_{i-2\hat{y}}-\bm{m}_{i+2\hat{y}}-2\bm{l}_i+\bm{l}_{i-2\hat{y}}-\bm{l}_{i+2\hat{y}}\Big)\label{Hl}
\end{eqnarray}
From these equations, we can find
\begin{eqnarray}
\frac{1}{2}\Big(\bm{B}^{\rm eff}_{m,i}+\bm{B}^{\rm eff}_{l,i}\Big)&=&\bm{B}^{\rm eff}_{j,2n},\ \frac{1}{2}\Big(\bm{B}^{\rm eff}_{m,i}-\bm{B}^{\rm eff}_{l,i}\Big)=\bm{B}^{\rm eff}_{j,2n+1}
\end{eqnarray}
where $i\equiv (j,2n)$ has been used. This clearly comes from the chain rules when transforming from derivatives of energy with respect to $\bm{M}_{j,k}$ to that with respect to $\bm{m}_i,\bm{l}_i$. The LLG equations for $\bm{m},\bm{l}$ then are derived in the exchange limit as, with $i\equiv (j,2n)$,
\begin{eqnarray}
\dot{\bm{m}}_{i}&=&\frac{1}{2}\Big(\dot{\bm{M}}_{j,2n}+\dot{\bm{M}}_{j,2n+1}\Big)=\frac{-\gamma}{2} (\bm{m}_i+\bm{l}_i)\times \frac{1}{2}(\bm{B}^{\rm eff}_{m,i}+ \bm{B}^{\rm eff}_{l,i})+\frac{-\gamma}{2} (\bm{m}_i-\bm{l}_i)\times \frac{1}{2}(\bm{B}^{\rm eff}_{m,i}-\bm{B}^{\rm eff}_{l,i})\nonumber\\
&&+\frac{\alpha}{2}(\bm{m}_i+\bm{l}_i)\times\Big(\dot{\bm{m}}_i+\dot{\bm{l}}_i\Big)+\frac{\alpha}{2}(\bm{m}_i-\bm{l}_i)\times\Big(\dot{\bm{m}}_i-\dot{\bm{l}}_i\Big)\approx\Big(\frac{1}{2}\gamma \bm{B}^{\rm eff}_{l,i}-\alpha \dot{\bm{l}_i}\Big)\times \bm{l}_i,\\
\dot{\bm{l}}_i&=&\frac{1}{2}\Big(\dot{\bm{M}}_{j,2n}-\dot{\bm{M}}_{j,2n+1}\Big)\approx \frac{1}{2}\gamma\bm{B}^{\rm eff}_{m,i}\times \bm{l}_i,
\end{eqnarray}
which are used in the discussions above.
\section{Thiele approach for DW velocity}
Using the Thiele approach~\cite{Thiele}, we can get exactly the same analytical DW velocity as Eq.~(\ref{DW_velocity}) as follows. After applying $\bm{M}\times$ on both sides of the LLGS equation, it reduces to
\begin{eqnarray}
0&=&\underbrace{\frac{-1}{M^2\gamma}\bm{M}\times\frac{d \bm{M}}{dt}}_{\text{Gyroscopic }\bm{B}^g}+\underbrace{\frac{-\alpha}{M\gamma}\frac{d\bm{M}}{dt}}_{\text{Dissipative }\bm{B}^\alpha}+\underbrace{\bm{B}^{\rm eff}}_{\text{Reversible } \bm{B}^r}+\underbrace{\frac{-(\bm{M}\cdot \bm{H}^{\rm eff})}{M^2}\bm{M}}_{\bm{B}^m}+\underbrace{\frac{-u}{M^2\gamma}\bm{M}\times \partial_x \bm{M}}_{\bm{B}^u}+\underbrace{\frac{-\beta u}{M\gamma}\partial_x \bm{M}}_{\bm{B}^\beta},
\end{eqnarray}
where the labellings for all terms follow from Thiele~\cite{Thiele} and we omitted the layer subindex $i=$U/L. The procedure is to calculate the force densities for all $\bm{B}^\mu$ terms, defined by $f^\mu_i\equiv - \bm{B}^\mu \cdot(\partial_i \bm{M})$, then integrate all of them over space. There is only $x$ and $t$ dependences of $\bm{M}$, thus we only need to deal with the nonzero $x$ components of the force densities, $0=\sum_\mu \int dx f^{\mu}_x$, where $\mu$ includes all the six terms above. Using the DW profiles and Hamiltonian defined in SI~Sec.~II, and the properties $\theta=\theta(x-\dot{q}t)$ and $\phi=\phi(t)$, explicit calculations show that
\begin{eqnarray}
f^g_{x}&=&\frac{-M}{\gamma}(\sin\theta)(\partial_x\theta)\dot{\phi},\ f^u_{x}=0,\ f^m_{x}=0,\ f^\alpha_{x}+f^\beta_{x}=\frac{-M(\alpha \dot{q}-\beta u)}{\gamma}(\partial_x\theta)^2,\ \int^\infty_{-\infty} dx f^r_{x}=2MB_{\rm SO},
\end{eqnarray}
after using $(\partial_t\theta) = -\dot{q}(\partial_x\theta)$. The total sum leads to
\begin{eqnarray}
0&=&\int dx\Big(f^g_x+f^\alpha_x+f^r_x+f^m_x+f^u_x+f^\beta_x\Big)\Rightarrow \dot{q}= u\frac{\beta}{\alpha}+\frac{B_{\rm SO}\gamma\Delta}{\alpha}+\frac{\Delta}{\alpha}\dot{\phi}.
\end{eqnarray}
We note that, this equation can actually be derived by combining Eq.~(\ref{dq_dt}) and Eq.~(\ref{phiU_eq}). Therefore, it gives the same result as that from the LLGS equation. This is very natural, since the Thiele approach is a procedure of manipulating the original three LLGS equations for each layer. Specifically, for the static state solution, $\dot{\phi}_{\rm U/L}=0$, this equation gives exactly the same DW velocity as in Eq.~(\ref{DW_velocity}). However, the Thiele approach cannot tell us the information about values of $\phi_{\rm U,L}$ as opposed to previous method [Eq.~(\ref{sin2phiU})], since in calculating $f^\mu_x$ we have rearranged the LLGS equation into a more restricted equation, which will lose some information contained in its original form. Specifically, LLGS equation contains three equations for $x,y,z$ components of $\partial_t \mathbf{M}$ for each upper and lower layers, but the Thiele equation reduces the number of equations to one for each layers, since we take the dot product $f^\mu_i\equiv - \bm{B}^\mu \cdot(\partial_i \bm{M})$, and only $i=x$ contributes due to the considered system geometry.